\documentclass{article}
\font\cero=cmss10 scaled 1728 
\usepackage{tensor}
\usepackage[english]{babel}
\usepackage{verbatim}
\usepackage{amsfonts}
\usepackage{graphicx}
\usepackage{amssymb}
\usepackage{amsmath}
\usepackage{braket}
\usepackage{comment}
\usepackage{float}
\usepackage{cite}
\usepackage{notoccite}
\usepackage[utf8]{inputenc}
\usepackage{hyperref}
\usepackage{appendix}
\usepackage{amsmath}
\DeclareMathOperator{\Det}{Det}
\DeclareMathOperator{\Tr}{Tr}

\usepackage[sort&compress,square,numbers]{natbib}

\begin{document}

\begin{flushleft}
{\cero A hypercomplex partition function for dissipative quantum field theory}\\

\end{flushleft} 
{\sf O. Cruz-Limón\(^1\), C. Ramirez-Romero\(^1\) and R. Cartas-Fuentevilla\(^2\).}\\
{\(^1\)\it Facultad de Ciencias Físico Matemáticas, Benemérita Universidad Autónoma de Puebla. P.O. Box 165, 72570 Puebla, México.}\\
{\(^2\)\it Instituto de F\'{\i}sica, Universidad Aut\'onoma de Puebla,
Apartado Postal J-48 72570 Puebla, Pue. M\'exico.}
\\ 
\noindent
oscar.cruzz@aol.com, \, cramirez@fcfm.buap.mx, \, rcartas@ifuap.buap.mx.\\

\noindent
ABSTRACT: We develop a finite-temperature formulation for a hypercomplex dissipative quantum field theory \cite{CartasOscar}, using the imaginary-time path-integral approach and the idempotent structure of the hypercomplex algebra. The resulting partition function naturally separates into two conjugate complex sectors whose recombination preserves the hypercomplex Hermitian structure while generating a nontrivial thermal phase. From this construction, the standard thermodynamic observables are obtained consistently, and the conventional relativistic charged Bose gas is recovered in the vanishing-dissipation limit. Beyond this equilibrium correspondence, the hypercomplex formulation reveals a distinctive perturbative hierarchy: dissipative effects first appear in the complementary phase sector, while corrections to ordinary real thermodynamic quantities arise only at higher order. These results show that dissipation can be encoded through an enlarged algebraic thermal structure without abandoning the familiar framework of relativistic finite-temperature field theory, opening a path toward broader applications of hypercomplex methods in dissipative, open, and effectively non-Hermitian quantum systems.\\


\section{Introduction}
\label{section 1}
Dissipation plays an important role in condensed matter physics, open quantum systems, quantum information, non-equilibrium statistical mechanics, and quantum field theory. Standard descriptions usually incorporate dissipative effects through nonself-adjoint Hamiltonians \cite{sergi}, open-system evolution equations \cite{open systems, dynamics in qft}, phenomenological damping terms \cite{amortiguamiento feno}, or third quantization techniques \cite{3ra cuantizacion}. Although these approaches successfully describe a wide range of physical regimes, establishing a direct connection between dissipation, field-theoretical dynamics, and thermodynamic structure remains a nontrivial problem \cite{filosofia en ciencia, disipacion inestabilidad}. An alternative possibility is to encode dissipative effects directly into an enlarged algebraic structure of the fields. Hypercomplex algebras provide a natural setting for this purpose because they introduce additional algebraic units and projection operators while retaining a unified field-theoretical description  \cite{metodos hipercomplejos}.

\vspace{0.2cm}
\noindent
In a previous work  \cite{CartasOscar}, we constructed a dissipative quantum field theory over a commutative hypercomplex ring containing the elliptic unit \(i\) and the hyperbolic unit \(j\). Within this framework, dissipative and interaction terms arise directly from the algebraic structure of the Lagrangian and Hamiltonian rather than from the introduction of an explicitly non-unitary evolution operator. The resulting theory admits well-defined normal modes, a real energy spectrum, and a consistent particle interpretation. An important feature of the construction is that the complete theory remains Hermitian with respect to the native hypercomplex conjugation, whereas complex effective parameters appear only after projection onto the ordinary \(\mathbb{C}\)-complex sectors. Thus, dissipation is associated with the enlarged algebraic structure rather than with a fundamental breakdown of Hermiticity.

\vspace{0.2cm}
\noindent
The purpose of the present work is to develop the finite-temperature sector of this hypercomplex dissipative theory. Using the imaginary-time path-integral formalism, we construct the grand-canonical partition function and show that the idempotent decomposition naturally produces two conjugate complex thermal sectors. Their recombination yields a Hermitian-valued hypercomplex partition function in which the dissipative parameter modifies both the quasiparticle scale and the thermal weights through a conjugate phase structure. From this partition function we derive the pressure, conserved-charge density, energy density, and entropy, and verify their thermodynamic consistency. The conventional relativistic charged Bose gas is recovered continuously when the dissipative deformation vanishes, while finite dissipation produces a characteristic separation between the real thermodynamic sector and the complementary (\(ij\)-hybrid) hypercomplex phase contribution.

\vspace{0.2cm}
\noindent
The paper is organized as follows. In Section \ref{seccion 2} we introduce the algebraic structure of the formalism. Section \ref{seccion 3} develops the grand-canonical partition function within the imaginary-time path-integral formalism, including the Gaussian integration prescription, the functional determinant, and its Matsubara representation. Section \ref{seccion 4} analyzes the resulting hypercomplex thermal structure and identifies the conjugate phase generated by the dissipative deformation. Section \ref{seccion 5} derives the corresponding thermodynamic observables and examines their internal consistency. Section \ref{seccion 6} establishes the continuous recovery of the conventional relativistic charged Bose gas in the vanishing-dissipation limit. Section \ref{seccion 7} explores the high- and low-temperature regimes, while Section \ref{seccion 8} examines the perturbative structure associated with weak dissipation. Section \ref{seccion 9} discusses the massless and gapless limits of the theory. Finally, Section \ref{conclusiones} summarizes the main results and outlines possible directions for future developments. Appendix \ref{apendice a} provides the detailed real-field representation of the projected Gaussian functional integrals and the mathematical justification of the integration-cycle prescription used in the main text.

\section{Hypercomplex field structure}
\label{seccion 2}
In this section, we establish the notation, define the algebraic structure of the theory and some of its properties. The hypercomplex ring \(\mathbb{H}\) is defined by,
\begin{eqnarray}
    \xi=x+iy+ju+ijv; \quad j^2=1; \quad i^2=-1; \quad (ij)^2=-1; \quad x,y,u,v \in \mathbb{R}. 
\label{def.num.hipercomplejos}
\end{eqnarray}
The conjugation of the complex units is as follows: for the hyperbolic complex unit \(\overline{j}=-j\), for the standard complex unit \(\overline{i}=-i\) and, for the hybrid complex unit \(\overline{ij}=ij\), in addition, for the latter we have \(ij=ji\). Thus, the modulus of the hypercomplex number $\xi$ in (\ref{def.num.hipercomplejos}) is,
\begin{equation}
    \xi \overline{\xi}= x^2 + y^2 - u^2 - v^2 + 2ij(xv-yu),
\label{modulo hipercomplejo}
\end{equation}
which is not a real number, instead it is a \textit{Hermitian} number, being the generalization to the hypercomplex scheme of a real number. This algebra extends the complex  numbers, while preserving associativity and commutativity. Expression (\ref{modulo hipercomplejo}) is invariant under the action of a compact times a noncompact group, \(U(1) \times SO(1,1)\). That is, under the circle rotations given by \(U(1)\), and under hyperbolic rotations of the connected component of the Lie group \(SO(1, 1)\) containing the group unity. 
With the hybrid complex unit \(ij\), we can define a Hermitian exponential \(e^{ij \theta}\), which can be expressed in the following form,
\begin{equation}
    e^{ij \theta} = \cos(\theta) + ij \sin(\theta).
\label{exponencial hibrida}
\end{equation}
 An expression invariant under the full action of the group \(U(1) \times SO(1,1)\) must take values in the Hermitian extension of the real numbers.

\subsection{Properties of the hypercomplex ring}
\label{seccion 2.1}
The idempotent basis of the ring \(\mathbb{H}\) has the elements,
\begin{eqnarray}
\begin{aligned}
    J^{+}&= \frac{1}{2}(1+j), \quad \quad (J^{+})^{n}=J^{+};\\
    J^{-}&= \frac{1}{2}(1-j), \quad \quad (J^{-})^{n}=J^{-}, \quad n=1,2,3,...;
\label{bases idempotentes}
\end{aligned}
\end{eqnarray}
with the properties,
\begin{equation}
    \begin{aligned}
        j J^{+}&=J^{+},    \hspace{1cm}       &  \overline{J^{+}}&=J^{-},  \hspace{1cm}   &  J^{+} + J^{-}&=1,\\
j J^{-}& = -J^{-},        &  \overline{J^{-}}&=J^{+},    &  J^{+} - J^{-}&= j,\\
J^{+} \cdot J^{-}&=0.      &         &             &  &
    \end{aligned}
\label{propiedades de H}
\end{equation}
The elements of the hypercomplex ring can be written in terms of the basis \(\{1,j\}\) or \(\{J^+,J^-\}\), as,
\begin{eqnarray}
\begin{aligned}
    \xi= z_1 + j z_2 &= (x_1 + ix_2) + j(y_1 + iy_2)\\
    &= \left[(x_1 + y_1) + i(x_2 + y_2) \right] \left(\frac{1+j}{2}\right) + \left[(x_1 - y_1) + i(x_2 - y_2) \right] \left(\frac{1-j}{2}\right)\\
    &=z^{+} J^{+} + z^{-}J^{-},
\label{numero hip. con bases idemp}
\end{aligned}
\end{eqnarray}
where \(z^{+}=(x_1 + y_1) + i(x_2 + y_2)\) and \(z^{-}=(x_1 - y_1) + i(x_2 - y_2)\). Throughout this work, we use the idempotent representation \(\{J^+, J^-\}\), particularly in the decomposition of the hypercomplex field in Eq. (\ref{Omega con bases idempotentes}). The following identities and properties of hypercomplex numbers will also be used \cite{Gurses, Silviu,CartasOscar,CartasAlex}:
\begin{enumerate}
        \item \((z_1 + jz_2)^{n}= J^{+} (z_1+z_2)^{n} + J^{-}(z_1 - z_2)^{n}\); \hspace{0.2cm} \(z_1,z_2 \in \mathbb{C}\),
        \label{propiedad 1}
        \item \(e^{J^- \chi}= J^{-} e^{\chi} + J^{+}\); \hspace{0.5cm} \(\chi \in \mathbb{C}\), 
        \label{propiedad 2}
        \item \(e^{j \chi}=e^{\chi}J^{+} + e^{-\chi}J^{-}\); \hspace{0.2cm} \(\chi \in \mathbb{C}\), 
        \label{propiedad 3}
        \item \(e^{J^+ \chi}= J^{+} e^{\chi} + J^{-}\);  \hspace{0.55cm} \(\chi \in \mathbb{C}\), 
         \label{propiedad 4}
        \item \(\ln(z_1+j z_2)=J^{+} \ln(z_1 + z_2) + J^{-}\ln(z_1 - z_2)\); \hspace{0.2cm} \(z_1,z_2 \in \mathbb{C}\).
        \label{propiedad 5}
\end{enumerate}
This algebraic extension makes it possible to formulate field theories where dissipative effects arise from the algebraic structure rather than depending on explicit non-unitary evolution operators.
Our model is not strictly a free-field theory, because interaction terms arise naturally from the invariant structure displayed in Eq. (\ref{modulo hipercomplejo}).

\section{Grand canonical partition function in the hypercomplex representation}
\label{seccion 3}
We begin from the standard definition of the partition function \cite{Kapusta, Alexander,Bellac,Landsman} in the context of \(U(1)\)-charged complex fields. In the construction of \( Z_{\mathbb{H}}\), the relevant exponentials involve both the hyperbolic unit, expressed either in the basis  \(\{1,j\} \) or \(\{J^{+}, J^{-}\}\), and the standard imaginary unit \(i\) associated with the \(U(1)\)-charged fields. The path-integral exponential therefore has a hybrid form,
\begin{equation}
e^{i(S+ijS')}=e^{iS}e^{-jS'}= J^{+} e^{iS-S'} + J^{-} e^{iS+S'},
\label{exponencial hibrida para Z}
\end{equation}
where the properties \ref{propiedad 2}-\ref{propiedad 4} listed in Sec. \ref{seccion 2.1} have been used. Further, the properties of the \(\mathbb{H}\)-ring allow us to manipulate these exponentials in such a way that the path integral required to calculate the partition function is completely complex, in the sense of the imaginary unit \(i\). The thermodynamic sector of our theory can be constructed directly from the hypercomplex Lagrangian introduced for the dissipative charged scalar field. In contrast with conventional approaches in which dissipation is incorporated phenomenologically into an already defined thermal theory, here both the dissipative deformation and the thermal structure originate from the same algebraic formulation. The idempotent decomposition of the hypercomplex field plays a central role because it separates the complete theory into two conjugate complex sectors while preserving the underlying hypercomplex structure. The starting point is the Lagrangian density introduced in \cite{CartasOscar} for two charged scalar fields,
\begin{equation}
     \mathcal{L}(\Omega, \overline{\Omega})=\frac{1}{2} \left [ \partial_{\mu}\Omega \cdot \partial^{\mu}\overline{\Omega} + \frac{\gamma}{2} (j\Omega \dot{\overline{\Omega}} + c.c)- m^2\Omega \overline{\Omega} \right ],  
\label{lagrangiana}
\end{equation}
where \(\Omega= \Phi + j \Psi\) is a hypercomplex field and \(\gamma\) is a dissipative parameter. Both fields \((\Phi, \Psi)\) are charged fields \(\Phi \rightarrow \phi_1 + i \phi_2\) and \(\Psi \rightarrow \psi_1 + i \psi_2\), thus,
\begin{equation}
    \Omega= \Phi + j \Psi= \phi_1 + i\phi_2 + j\psi_1 + ij\psi_2; \hspace{2.3cm} \phi_{1,2}, \psi_{1,2} \in \mathbb{R}.
\label{omega 4 componentes}
\end{equation}
This Lagrangian (\ref{lagrangiana}) represents the entire system; in addition, the field \(\Phi\) represents the system of interest, and the field \(\Psi\) represents the environment.  We can now observe that the structure of the elements in (\ref{lagrangiana}) has the form of the modulus of a hypercomplex number (\ref{modulo hipercomplejo}).  The hypercomplex field \(\Omega\) can be rewritten using the idempotent basis (\ref{bases idempotentes}) and the property in (\ref{numero hip. con bases idemp}), resulting,
\begin{eqnarray}
    \Omega= J^{+} \Omega^{+} + J^{-} \Omega^{-}; \hspace{1.5cm} \quad \Omega^{+}=(\Phi+\Psi),
    \quad
    \Omega^{-}=(\Phi - \Psi).
\label{Omega con bases idempotentes}
\end{eqnarray}
where \(\Omega^{+}\) and \(\Omega^{-}\) are standard complex fields. The hypercomplex conjugation interchanges the idempotent projectors, and therefore, 
\begin{equation}
    \overline{\Omega} = J^{-} \overline{\Omega^{+}} + J^{+} \overline{\Omega^{-}},
\label{campo hipercomplejo conjugado}
\end{equation}
thus,
\begin{equation}
    \Omega \overline{\Omega} = J^{+} \Omega^{+} \overline{\Omega^{-}} + J^{-} \Omega^{-} \overline{\Omega^{+}},
\label{omega como modulo}
\end{equation}
and the complete Lagrangian (\ref{lagrangiana}) decomposes as,
\begin{equation}
    \mathcal{L} = J^{+} \mathcal{L}_{+} + J^{-} \mathcal{L}_{-}.
\label{lagrangiana descompuesta en idempotentes}
\end{equation}
The two projected Lagrangian densities are,
\begin{equation}
\begin{aligned}
    \mathcal{L}_{+} = \frac{1}{2} \left[\partial_{\mu} \Omega^{+} \partial^{\mu} \overline{\Omega^{-}} + \frac{\gamma}{2} \left(\Omega^{+} \dot{\overline{\Omega^{-}}} - \overline{\Omega^{-}} \dot{\Omega}^{+} \right) - m^2 \Omega^{+} \overline{\Omega^{-}} \right],\\
    \mathcal{L}_{-} = \frac{1}{2} \left[\partial_{\mu} \Omega^{-} \partial^{\mu} \overline{\Omega^{+}} - \frac{\gamma}{2} \left(\Omega^{-} \dot{\overline{\Omega^{+}}} - \overline{\Omega^{+}} \dot{\Omega}^{-} \right) - m^2 \Omega^{-} \overline{\Omega^{+}} \right].
\label{lagrangianas en terminos de bases idempotentes}
\end{aligned}
\end{equation}
The opposite signs of the first-order time derivative are a direct consequence of the action of the hyperbolic unit on the idempotent sectors. Thus, the two projected fields are not independent copies of the same dynamics; rather, they represent conjugate components of the complete hypercomplex system. For convenience, both sectors can be represented simultaneously by introducing \(s = \pm 1\) and defining,
\begin{equation}
\begin{matrix}
X_{+} = \Omega^{+}, \hspace{1cm} & Y_{+} = \overline{\Omega^{-}},  \\
X_{-} = \Omega^{-}, \hspace{1cm} & Y_{-} = \overline{\Omega^{+}}. \\
\end{matrix}
\label{definicion de campos complejos en terminos de XY}
\end{equation}
Then,
\begin{equation}
    \mathcal{L}_{s} = \frac{1}{2} \left[ \partial_{\mu} X_{s} \partial^{\mu} Y_{s} + s \frac{\gamma}{2} \left( X_{s} \dot{Y}_{s} - Y_{s} \dot{X}_{s} \right) - m^2 X_{s}  Y_{s}   \right].
\label{lagrangiana en terminos de XY}
\end{equation}
This expression will serve as the starting point for constructing the partition function \(Z_{\mathbb{H}}\).

\subsection{Canonical structure}
The canonical momenta derived directly from Eq. (\ref{lagrangiana en terminos de XY}) are,
\begin{equation}
    \pi^{(s)}_{X} = \frac{\partial \mathcal{L}_{s}}{\partial \dot{X}_{s}} = \frac{1}{2} \left(\dot{Y}_{s} - s \frac{\gamma}{2} Y_{s} \right); \hspace{1.5cm} \pi^{(s)}_{Y} = \frac{\partial \mathcal{L}_{s}}{\partial \dot{Y}_{s}} = \frac{1}{2} \left(\dot{X}_{s} + s \frac{\gamma}{2} X_{s} \right),
\label{momentos canonicos en termimos de XY}
\end{equation}
therefore,
\begin{equation}
    \dot{X}_{s} = 2 \pi^{(s)}_{Y} - s \frac{\gamma}{2} X_{s}; \hspace{1.5cm} \dot{Y}_{s} = 2 \pi^{(s)}_{X} + s \frac{\gamma}{2} Y_{s}.
\label{momentos X e Y}
\end{equation}
The Legendre transformation yields,
\begin{equation}
    \mathcal{H}_{s} = \pi^{(s)}_{X} \dot{X}_{s} + \pi^{(s)}_{Y} \dot{Y}_{s} - \mathcal{L}_{s}, 
\label{transf. de Legendre 1}
\end{equation}
which gives, 
\begin{equation}
    \mathcal{H}_{s} = 2 \pi^{(s)}_{X} \pi^{(s)}_{Y} + s \frac{\gamma}{2} \left( Y_{s} \pi^{(s)}_{Y} - X_{s} \pi^{(s)}_{X} \right) + \frac{1}{2} \nabla X_{s} \cdot \nabla Y_{s} + \frac{1}{2} m^{2}_{\texttt{eff}} X_{s} Y_{s}; 
\label{hamiltoniano en terminos de XY}
\end{equation}
where the effective mass is defined by,
\begin{eqnarray}
    m^{2}_{\texttt{eff}} = m^2 - \frac{\gamma^{2}}{4}.
\label{masa efectiva}
\end{eqnarray}
The same parameter controls the oscillatory part of the classical dispersion relation. In particular, stability of the quasiparticle spectrum requires,
\begin{equation}
     m^{2}_{\texttt{eff}} > 0. 
\end{equation}
The Euler-Lagrange equations derived directly from Eq. (\ref{lagrangiana en terminos de XY}) are,
\begin{equation}
    \left(\partial_{\mu} \partial^{\mu} + s \gamma \partial_{t} + m^2 \right) X_{s} = 0, 
\label{ecuaciones de euler lagrange en terminos de X}
\end{equation}
showing explicitly that the \(J^{+}\) and \(J^{-}\) sectors carry opposite dissipative evolution while sharing the same effective spectral scale. For a more detailed discussion of the solution to these equations of motion (\ref{ecuaciones de euler lagrange en terminos de X}) and their implications for the present theory, see \cite{CartasOscar}. It is worth emphasizing that the condition \(\Omega^{-}=0\) may consistently be imposed at the level of the classical solutions \cite{RamirezNeph}, after the equations of motion have been derived from the complete hypercomplex action. Since the equation governing \(\Omega^{-}\) is linear and homogeneous, initial conditions \(\Omega^{-}(t_{0}) = \dot{\Omega}^{-}(t_{0}) =0\) select the dynamically damped sector without generating the exponentially growing classical solution. However, \(\Omega^{-}\) cannot be imposed as a truncation of the action or of the functional integral itself. Indeed, because hypercomplex conjugation exchanges the idempotent sectors and \(J^{+} \cdot J^{-}=0\), setting \(\Omega^{-}=0\) also implies \(\overline{\Omega^{-}} = 0\), causing the bilinear terms that define the projected action to vanish and rendering the functional theory degenerate. Therefore, the \(J^{-}\) sector must be retained as the conjugate algebraic degree of freedom required for the variational formulation and for the construction of the non-degenerate partition function, even when the physically selected classical solutions are restricted to the damped sector.

\subsection{Conserved charge and grand canonical Hamiltonian}
Because the fields are charged under the \(U(1)\)-symmetry, the infinitesimal transformation can be written as,
\begin{equation}
    X_{s} \rightarrow X_{s} + i \alpha X_{s},  \hspace{1.5cm}  Y_{s} \rightarrow Y_{s} - i \alpha Y_{s}.
\label{transformacion infinitesimal}
\end{equation}
The corresponding charge density is
\begin{equation}
    \rho_{s} = i \left( X_{s} \pi^{(s)}_{X} - Y_{s} \pi^{(s)}_{Y} \right).
\label{densidad de carga en terminos de XY}
\end{equation}
The factor \(i\) is essential for interpreting \(\mu\) as the usual physical chemical potential. Defining,
\begin{equation}
    T_{s} \equiv Y_{s} \pi^{(s)}_{Y} - X_{s} \pi^{(s)}_{X}; 
\end{equation}
one obtains,
\begin{equation}
    \rho_{s} = - i T_{s}. 
\end{equation}
The grand canonical Hamiltonian is therefore,
\begin{equation}
    K_{s} = H_{s} - \mu Q_{s},
\label{grand Canonical Hamiltoniano}
\end{equation}
or locally,
\begin{equation}
    \mathcal{K}_{s} = \mathcal{H}_{s} - \mu \rho_{s}.
\label{grand canonical LOCAL hamiltoniano}
\end{equation}
Using the expressions (\ref{hamiltoniano en terminos de XY}) and (\ref{densidad de carga en terminos de XY}), we obtain,
\begin{equation}
    \mathcal{K}_{s} = 2 \pi_{X} \pi_{Y} + \left( s \frac{\gamma}{2} + i \mu \right) T_{s} + \frac{1}{2} \nabla X_{s} \cdot \nabla Y_{s} + \frac{1}{2} m^{2}_{\texttt{eff}} X_{s} Y_{s}. 
\label{grand canonical LOCAL hamiltoniano 2}
\end{equation}
The use of the grand canonical ensemble follows the standard construction for charged bosonic fields at finite temperature \cite{Kapusta,Alexander,Bellac,Landsman, integral de trayectoria 1,integral de trayectoria 2, dense matter}. In the present formulation, the dissipative contribution naturally accompanies the chemical potential as a consequence of the hypercomplex algebra.

\subsection{Imaginary-time functional integral}
The grand canonical partition function of each idempotent sector is
\begin{equation}
    Z_{s} = \texttt{Tr} \exp \left[- \beta \left(H_{s} - \mu Q_{s} \right) \right]. 
\label{z en terminos de la traza}
\end{equation}
The corresponding phase-space functional representation is \cite{Kapusta,Alexander,Bellac,Landsman, integral de trayectoria 1,integral de trayectoria 2, dense matter},
\begin{equation}
    Z_{s} = \int \mathcal{D} \pi_{X} \mathcal{D} \pi_{Y} \mathcal{D} X \mathcal{D} Y \exp \left\{\int_{0}^{\beta} d\tau \int d^3 x \left[i \pi_{X} \partial_{\tau} X + i \pi_{Y} \partial_{\tau} Y - \mathcal{K}_{s} \right] \right\};
\label{z en terminos de representacion funcional}
\end{equation}
with bosonic periodicity,
\begin{equation}
    X_{s} \left(\tau + \beta, \mathbf{x} \right) = X_{s} (\tau, \textbf{x}),  \hspace{1.5cm} Y_{s}\left(\tau + \beta, \textbf{x} \right) = Y_{s}(\tau, \textbf{x}).
\label{periodicidad bosonica}
\end{equation}
For convenience we introduce,
\begin{equation}
    c_{s} = -s \frac{\gamma}{2} - i \mu. 
\label{cs}
\end{equation}
The momentum-dependent part of the exponent becomes,
\begin{equation}
    -2 \pi_{X} \pi_{Y} + \pi_{X} \left(i \dot{X} - c_{s} X \right) + \pi_{Y} \left(i \dot{Y} + c_{s} Y \right).
\label{momentos dependientes}
\end{equation}
The algebraic identity,
\begin{equation}
    -2 pq + Ap + Bq = -2 \left(p - \frac{B}{2}\right) \left(q - \frac{A}{2} \right) + \frac{1}{2} AB,
\label{identidad algebraica 1}
\end{equation}
allows the canonical momenta to be integrated exactly. The remaining field-dependent contribution is,
\begin{equation}
    \frac{1}{2} \left(i \dot{X} - c_{s} X \right) \left(i \dot{Y} + c_{s} Y \right).
\label{contribucion de campo remanente}
\end{equation}
The decomposition into real field components establishes explicitly that the projected functional integral is quadratic, and therefore Gaussian, on the underlying real field space. Positive definiteness, however, is a separate issue associated with the convergence of the integration cycle. Since the projected idempotent actions are not generically positive definite on the real contour, the corresponding Gaussian integrals are defined by analytic continuation to appropriate steepest-descent cycles, as discussed in Appendix \ref{apendice a}. We now define the sector-dependent complex chemical potentials,
\begin{equation}
    \mu_{s} = \mu - is \frac{\gamma}{2},
\label{potencial quimico modificado por la disipacion}
\end{equation}
or explicitly,
\begin{equation}
    \mu_{+} = \mu - i \frac{\gamma}{2} , \hspace{1cm} \mu_{-} = \mu + i \frac{\gamma}{2}.
\label{ 2 potenciales quimicos modificados}
\end{equation}
For real \(\mu\) and \(\gamma\), we have,
\begin{equation}
    \mu_{-} = \mu_{+}^{\ast}.
\label{potenciales quimicos conjudagos}
\end{equation}
After integrating by parts and using the periodic boundary conditions, the Euclidean action can be written as,
\begin{equation}
    S_{E,s} = \frac{1}{2} \int_{0}^{\beta} d\tau \int d^{3}x Y_{s} \mathcal{D}_{s} X_{s}, 
\label{accion euclidea s}
\end{equation}
where
\begin{equation}
    \mathcal{D}_{s} = - \left(\partial_{\tau} + \mu_{s} \right)^{2} - \nabla^2 + m^{2}_{\texttt{eff}}.
\label{operador D definicion}
\end{equation}

\subsection{Gaussian structure and integration-cycle prescription}
\label{subseccion 3.4}
Before evaluating the functional determinant, it is important to specify the meaning of the Gaussian integration appearing in the projected idempotent sectors. Although the Euclidean action in Eq. (\ref{accion euclidea s}) is quadratic in the fields, its structure differs from the conventional positive-definite complex Gaussian. In particular, each projected action couples two conjugate components belonging to opposite idempotent sectors and has the generic bilinear form,
\begin{equation}
    S_{E,s} \sim \int_{x} Y_{s}\mathcal{D}_{s}X_s.
\label{3.4.1}
\end{equation}
Decomposing \(X_{s}\) and \(Y_{s}\) into their real and imaginary components shows that the functional integral is indeed Gaussian over the underlying real field space. However, the corresponding quadratic form is generically off-diagonal and is not positive definite on the naive real integration contour. Therefore, the Gaussian character of the functional integral and its convergence must be regarded as two distinct issues. A detailed real-component representation of this structure is given in Appendix \ref{apendice a}. The projected functional integral is defined by analytic continuation onto an appropriate complex integration cycle. More specifically, we employ a steepest-descent prescription, in which the integration variables are complexified and the cycle is chosen such that the real part of the Euclidean action increases toward \(+ \infty\) along its asymptotic directions. Thus,
\begin{equation}
    \text{Re} \, S_{E} \longrightarrow +\infty
    \qquad
    \Longrightarrow\qquad
    | e^{-S_{E}}|\longrightarrow 0,
\label{3.4.2}
\end{equation}
which provides exponential damping of the integrand. In its general formulation, this procedure is related to Morse and Picard–Lefschetz theory, where convergent middle-dimensional integration cycles are associated with critical points of the complexified action \cite{witten,cristoforetti,cristoforetti 2}. For the quadratic theory considered here, the prescription is particularly transparent. We first define the functional integral at finite spatial volume and with an ultraviolet regulator, such that only a finite number of Fourier-Matsubara modes is retained. The functional integral then reduces to a finite product of ordinary complex Gaussian integrals. After diagonalization, each non-degenerate mode can be represented schematically as,
\begin{equation}
    S_{\alpha} = \lambda_{\alpha} x_{\alpha} y_{\alpha},
\label{3.4.3}
\end{equation}
where \(\alpha\) denotes collectively the sector and mode labels. Introducing,
\begin{equation}
    r_{\alpha} = \frac{x_{\alpha} + y_{\alpha}}{\sqrt{2}},
    \hspace{1cm}
    q_{\alpha} = \frac{x_{\alpha} - y_{\alpha}}{\sqrt{2}},
\label{3.4.4}
\end{equation}
one obtains,
\begin{equation*}
    S_{\alpha} = \frac{\lambda_{\alpha}}{2} \left(r^{2}_\alpha - q^2_{\alpha} \right).
\label{3.4.5}
\end{equation*}
Thus, the naive real contour contains one damped and one growing Gaussian direction and is not, in general, a convergent Euclidean integration cycle. For a nonzero eigenvalue written as,
\begin{equation}
    \lambda_{\alpha} = |\lambda_{\alpha}| e^{i \vartheta_\alpha},
\label{3.4.6}
\end{equation}
a convergent cycle can instead be chosen through the rotations,
\begin{equation}
    r_{\alpha} = e^{-i \frac{\vartheta_{\alpha}}{2} } u_{\alpha},
    \hspace{1cm}
    q_{\alpha} = i e^{-i \frac{\vartheta_{\alpha}}{2} } v_{\alpha};
    \hspace{1cm}
    u_\alpha, v_{\alpha} \in \mathbb{R}.
\label{3.4.7}
\end{equation}
On this cycle the quadratic action becomes,
\begin{equation}
    S_{\alpha} = \frac{|\lambda_{\alpha}|}{2} \left( u^2_{\alpha} + v^{2}_{\alpha} \right).
\label{3.4.8}
\end{equation}
This explicit construction is the reason why the steepest-descent prescription can be applied directly in the present theory. Since the action is quadratic, the relevant cycles can be obtained mode by mode without requiring the full machinery associated with interacting theories containing several competing saddle points or Stokes phenomena. The two idempotent sectors are related by hypercomplex conjugation, and their integration cycles are correspondingly chosen as conjugate cycles: \(\mathcal{C}_{-} = \mathcal{C}_{+}^{\ast}\), thus, the projected functional integrals preserve the relation \(Z_{-} = Z_{+}^{\ast}\). The contour prescription also preserves the phase associated with the complex Gaussian determinant. While the positive Gaussian obtained after the rotation produces the usual inverse-modulus contribution, the complex Jacobian and orientation of the integration cycle supply the corresponding phase. Therefore, each non-degenerate complex bosonic mode produces, up to field-independent normalization,
\begin{equation*}
    Z_{\alpha} \propto \lambda_{\alpha}^{-1},
\end{equation*}
instead of simply \(|\lambda_{\alpha}|^{-1}\). This establishes the structure of the determinant used below. It is important to distinguish, however, the convergence of each Gaussian mode from the definition of the complete functional determinant. The condition \(\lambda_{\alpha} \neq 0\) guarantees that an individual quadratic saddle is non-degenerate and that its Gaussian integral can be defined on a convergent steepest-descent cycle. It does not imply convergence of the infinite product over all field modes. Therefore, the complete determinant is understood as a regularized functional determinant. Possible zero eigenvalues, if encountered, must be separated from the Gaussian fluctuation determinant and treated independently rather than simply discarded. The explicit real-field construction, the detailed steepest-descent analysis, and the treatment of the associated functional determinant are presented in Appendix \ref{apendice a}. With this prescription established, we now turn to the Fourier-Matsubara representation of \(\mathcal{D}_{s}\) and to the evaluation of its spectrum.


\subsection{Matsubara representation and functional determinant}
\label{seccion 3.5}
The periodic fields are expanded in bosonic Matsubara modes,
\begin{equation}
    X_{s} (\tau, \textbf{x}) = \sqrt{\frac{\beta}{V}} \sum_{n, \textbf{p}} e^{i(\textbf{p} \cdot \textbf{x} + \omega_{n} \tau)} X_{s,n} (\textbf{p}),
\label{modos de matsubara bosonicos}
\end{equation}
and analogously for \(Y_{s}\), where
\begin{equation}
    \omega_{n} = 2 \pi n T, \hspace{1.5cm} n \in \mathbb{Z}.
\label{omega 2 Pi n T}
\end{equation}
The normalization in (\ref{modos de matsubara bosonicos}) is chosen such that each Fourier amplitude is dimensionless. Defining energy as,
\begin{equation}
    E^{2}_{\textbf{p}} = \textbf{p}^2 + m^2_{\texttt{eff}},
\label{energia modificada por la disipacion}
\end{equation}
the eigenvalues of the Euclidean operators are,
\begin{equation}
    \lambda_{s} (n, \textbf{p}) = \left(\omega_{n} - i\mu_{s} \right)^2 + E^2_{\textbf{p}}
\label{eigenvalores del operador euclidino D}
\end{equation}
or explicitly,
\begin{equation}
    \lambda_{+} = \left( \omega_{n} -\frac{\gamma}{2} - i \mu \right)^2 + E^2_{\textbf{p}}, \hspace{1.5cm} \lambda_{-} = \left( \omega_{n} +\frac{\gamma}{2} - i \mu \right)^2 + E^2_{\textbf{p}}.
\label{eigenvalores del operador euclidino D EXPLICITOS}
\end{equation}
The two spectra satisfy the conjugation relation,
\begin{equation}
    \lambda_{-} (-n, \textbf{p}) = \lambda_{+} (n, \textbf{p})^{\ast}.
\label{relacion de conjugacion de eigenvalores}
\end{equation}
The spectral domain relevant for the thermal construction can now be stated explicitly. From Eq. (\ref{eigenvalores del operador euclidino D EXPLICITOS}), the real part of the eigenvalues is, 
\begin{equation}
    \text{Re} \, \lambda_{s} (n, \textbf{p}) = \left(\omega_{n} - s \frac{\gamma}{2} \right)^2 + E^2_{\textbf{p}} - \mu^2, 
    \hspace{1.5cm}
    s= \pm. 
\label{parte real de los valores propios de Ds}
\end{equation}
In the stable non-condensed thermal region,
\begin{equation}
    m^2_{\texttt{eff}} > 0, 
    \hspace{1.5cm}
    |\mu| < m_{\texttt{eff}}.
\label{region termal estable sin condensacion}
\end{equation}
Moreover,
\begin{equation*}
    E^2_{\texttt{p}} = \textbf{p}^2 +  m^2_{\texttt{eff}} \geq m^2_{\texttt{eff}} 
    \hspace{1cm}
    \Longrightarrow
    \hspace{1cm}
    E^2_{\texttt{p}} - \mu^2 \geq \, m^2_{\texttt{eff}} - \mu^2 > 0.
\end{equation*}
Therefore, 
\begin{equation}
    \text{Re} \, \lambda_{s} (n, \textbf{p}) > 0,
\label{valores propios reales}
\end{equation}
for every Matsubara frequency and momentum considered in the thermal sector. Hence no eigenvalue crosses the origin within this parameter domain \(\lambda_{s}(n, \mathbf{p}) \neq 0\), and all Gaussian fluctuation modes remain non-degenerate. Moreover, the spectrum lies entirely in the open right half of the complex plane, allowing the phases and the corresponding steepest-descent integration cycles introduced in Sec. \ref{subseccion 3.4} to be chosen continuously throughout the thermal domain. We therefore fix the principal branch of the complex logarithm
\begin{equation*}
    \ln z = \ln |z| + i \texttt{Arg} z, 
    \hspace{1.5cm}
    -\pi < \texttt{Arg} z \leq \pi.
\end{equation*}
For the present spectrum\footnote{Throughout the remainder of the paper, logarithms of complex projected quantities are understood on the principal branch defined above.} this reduces to,
\begin{equation*}
    -\frac{\pi}{2} < \texttt{Arg} \lambda_{s} (n, \textbf{p}) < \frac{\pi}{2},
\end{equation*}
because \(\text{Re}\, \lambda_{s} (n , \textbf{p}) > 0\). Hence, no eigenvalue crosses the origin within this parameter domain; equivalently, \(\lambda_{s} \neq 0\), and all Gaussian fluctuation modes remain nondegenerate. Moreover, using Eq. (\ref{relacion de conjugacion de eigenvalores}),
\begin{equation*}
    \ln \lambda_{-} (-n, \textbf{p}) = \left[\ln \lambda_{+} (n, \textbf{p}) \right]^{\ast}.
\end{equation*}
Therefore, a regularization preserving the Matsubara conjugation pairing also preserves,
\begin{equation*}
    \Tr_{\texttt{reg}} \ln \mathcal{D}_{-} = \left(\Tr_{\texttt{reg}} \ln \mathcal{D}_{+} \right)^{\ast}.
\end{equation*}
This condition guarantees the mode-by-mode applicability of the Gaussian contour prescription; the remaining infinite product over modes is understood separately through the regularized functional determinant. Therefore, after summation over the complete Matsubara spectrum, the projected partition functions are complex conjugates,
\begin{equation}
    Z_{-} = Z_{+}^{\ast}.
\label{funciones de particion conjugadas 1}
\end{equation}
This result is also obtained directly from the real Gaussian formulation described in Appendix \ref{apendice a}, where the two projected integration cycles are chosen as conjugate steepest-descent cycles. For a complex bosonic field, Gaussian functional integration yields\footnote{Formally, \(\Det \mathcal{D}_{s} \equiv \prod_{n, \textbf{p}} \, \lambda_{s} (n, \textbf{p})\); throughout the paper, this infinite product is understood through the regularized functional determinant defined below. } \cite{Kapusta,Alexander,Bellac,Landsman, integral de trayectoria 1,integral de trayectoria 2, dense matter}:
\begin{equation}
    Z_{s} \propto \left[ \Det \mathcal{D}_{s} \right]^{-1},
\label{integral gaussiana para campos complejos 1}
\end{equation}
and hence, 
\begin{equation}
    \ln Z_{s} = -\Tr_{\texttt{reg}} \ln \mathcal{D}_{s} + \text{constant}.
\label{logarimo de Zs esquematicamente}
\end{equation}
The complete hypercomplex partition function is reconstructed as,
\begin{equation}
    Z_{\mathbb{H}} = J^{+} Z_{+} + J^{-} Z_{-},
\label{logaritmo de Z-H en terminos de las bases idempotentes}
\end{equation}
and therefore,
\begin{equation}
\begin{aligned}
    \ln Z_{\mathbb{H}} &= J^{+} \left[\ln Z_{0,+} - \sum_{(n, \mathbf{p}) \neq (0, \mathbf{0})} \ln \left\{\beta^2 \left[ (\omega_{n} - i \mu_{+})^2 + E^2_{\textbf{p}} \right]\right\} \right] \\
    &+ J^{-} \left[\ln Z_{0,-} - \sum_{(n, \textbf{p}) \neq (0, \mathbf{0})} \ln \left\{\beta^2 \left[ (\omega_{n} - i \mu_{-})^2 + E^2_{\textbf{p}} \right]\right\} \right].
\label{ln Z-H con sumas de matsubara 1}
\end{aligned}
\end{equation}
The quantities \(Z_{0, \pm}\) denote the contributions associated with the static Fourier sector \((n,\textbf{p}) = (0,\mathbf{0})\). The static Fourier mode should not be confused with a zero eigenvalue of \(\mathcal{D}_s\). In the thermal domain considered here, all eigenvalues remain nonzero. Their detailed role in the condensate sector is not required for the thermal fluctuation analysis developed here. The separation of the static Fourier mode is therefore a bookkeeping choice rather than the removal of a singular Gaussian mode. Indeed, from Eq. (\ref{eigenvalores del operador euclidino D EXPLICITOS}), 
\begin{equation*}
    \lambda_{s} (0, \mathbf{0}) = m^2 - \mu^2 + i s \mu \gamma,
\end{equation*}
which remains nonzero throughout the thermal domain (\ref{region termal estable sin condensacion}). Its possible interpretation as a background or macroscopically occupied mode is left open and should not be identified with a true zero eigenvalue of \(\mathcal{D}_s\). In the strict non-condensed domain considered here, the isolated static Fourier contribution is also subextensive. Indeed, as long as \(\lambda_{s} (0, \mathbf{0}) \neq 0\) and the mode is not macroscopically occupied, \(\ln Z_{0,s}\) is an \(O (V^{0})\) contribution, whereas the bulk thermal part of \(\ln Z_{s}\) scales as \(O(V)\). Consequently, the contribution of the isolated static mode to intensive thermodynamic quantities, such as the pressure, charge density, energy density, and entropy density, is suppressed as \(O(V^{-1})\) and vanishes in the thermodynamic limit. We nevertheless keep \(Z_{0,s}\) formally separated because this argument need not apply if the zero-momentum sector becomes macroscopically occupied or approaches a genuinely degenerate configuration.

\subsubsection{Evaluation of the Matsubara sums}
Consider the generic sum,
\begin{equation}
    \mathcal{S}(E, \mu_{c}) = \sum_{n = - \infty}^{+\infty} \ln \left[ (\omega_{n} - i \mu_{c})^2 + E^2 \right],
\label{suma de matsubara generica}
\end{equation}
where \(\mu_{c}\) may be complex. Rather than summing the logarithm directly, we differentiate with respect to \(E\),
\begin{equation}
    \frac{\partial \mathcal{S}}{\partial E}  = 2 E \sum_{n} \frac{1}{(\omega_{n} -i \mu_{c} )^2 + E^2}.
\label{derivada de E para suma de matsubara}
\end{equation}
The standard bosonic Matsubara sum, analytically continued to complex \(\mu\), gives \cite{Kapusta,Alexander,Bellac,Landsman,dense matter},
\begin{equation}
    \sum_{n} \frac{1}{(\omega_{n} -i \mu_{c} )^2 + E^2} =  \frac{\beta}{2E} \left[1 + n_{B} (E - \mu_{c}) + n_{B} (E + \mu_{c}) \right],
\label{evaluacion de suma de matsubara generica}
\end{equation}
where
\begin{eqnarray}
    n_{B} (z) = \frac{1}{e^{\beta z} - 1},
\label{distribucion de Bose Einstein}
\end{eqnarray}
is the Bose-Einstein distribution. It follows that,
\begin{equation}
    \frac{\partial \mathcal{S}}{\partial E} = \beta \left[ 1 + n_{B} (E - \mu_{c}) + n_{B} (E + \mu_{c}) \right].
\label{derivada de E de la suma generica de Matsu}
\end{equation}
Now, integrating (\ref{derivada de E de la suma generica de Matsu}) with respect to \(E\), we obtain,
\begin{equation}
    \mathcal{S}(E, \mu_{c}) = \beta E + \ln \left( 1 - e^{-\beta(E - \mu_{c})} \right) + \ln \left( 1 - e^{-\beta(E + \mu_{c})} \right) + C,
\label{integracion de E en la suma de matsubara generica}
\end{equation}
where \(C\) is fixed by a thermodynamic-parameter-independent normalization of the functional measure and can therefore be taken independent of \(\beta\) and \(\mu_c\). Applying Eq. (\ref{integracion de E en la suma de matsubara generica}) to the two projected sectors (\ref{ln Z-H con sumas de matsubara 1}), and passing to the continuum limit,
\begin{equation*}
    \sum_{\textbf{p}} \rightarrow V \int \frac{d^3 p}{(2 \pi)^3},
\end{equation*}
one obtains, 
\begin{equation}
\begin{aligned}
    \ln Z_{\mathbb{H}} &= J^{+} \left\{ \ln Z_{0,+} - V \int \frac{d^3 p}{(2 \pi)^3} \left[\ln \left(1 - e^{-\beta(E_{\textbf{p}} - \mu_{+} )} \right) + \ln \left(1 - e^{-\beta(E_{\textbf{p}} + \mu_{+} )} \right) \right]  \right\}\\
    &+ J^{-} \left\{ \ln Z_{0,-} - V \int \frac{d^3 p}{(2 \pi)^3} \left[\ln \left(1 - e^{-\beta(E_{\textbf{p}} - \mu_{-} )} \right) + \ln \left(1 - e^{-\beta(E_{\textbf{p}} + \mu_{-} )} \right) \right]  \right\}\\
    &- \beta V \int \frac{d^3 p }{(2 \pi)^3} E_{\textbf{p}}. 
\label{Z-H en terminos de logaritmos con densidad de energia de V}
\end{aligned}
\end{equation}
Eq. (\ref{Z-H en terminos de logaritmos con densidad de energia de V}) displays separately the zero-point contribution and the finite-temperature contributions. The thermal factors possess the same particle-antiparticle structure as in the relativistic charged Bose gas, while the hypercomplex deformation appears entirely through the conjugate pair \(\mu_\pm\) (\ref{potencial quimico modificado por la disipacion}) and through \(m^2_{\texttt{eff}}\) (\ref{masa efectiva}).

\section{Hypercomplex thermal structure}
\label{seccion 4}
To make the hypercomplex contribution explicit, define,
\begin{equation}
    \theta_{T} \equiv \frac{\beta \gamma}{2},
\label{contribucion termica}
\end{equation}
and,
\begin{equation}
    r_{-}(p) = e^{-\beta(E_{\textbf{p}} - \mu )}, \hspace{1.5cm} r_{+}(p) = e^{-\beta(E_{\textbf{p}} + \mu )}.
\label{definicion de r_{+ -}}
\end{equation}
The thermal factors then satisfy,
\begin{equation}
    e^{-\beta (E_{\textbf{p}} - \mu_{+})} = r_{-} e^{-i \theta_{T}} ,  \hspace{1.5cm}  e^{-\beta (E_{\textbf{p}} + \mu_{+})} = r_{+} e^{+i \theta_{T}};
\label{factores termicos}
\end{equation}
whereas the \(J^{-}\) sector carries their complex conjugates.
This structure shows that the dissipative parameter does not act as a conventional real shift of the chemical potential. Instead, it introduces a thermal phase,
\begin{equation}
    e^{\pm i \theta_{T}} = e^{\pm i \beta \gamma/ 2},
\label{definicion de FASE TERMICA}
\end{equation}
multiplying the ordinary Boltzmann weights. The thermal sector therefore contains simultaneously an equilibrium statistical suppression determined by \(E_{\textbf{p}} \mp \mu\) and an oscillatory contribution generated by the hypercomplex deformation. For later convenience, the real part of each logarithm in (\ref{Z-H en terminos de logaritmos con densidad de energia de V}) may be written as,
\begin{equation}
    \Re \ln \left(1 - r e^{\pm i \theta_{T}} \right) = \frac{1}{2} \ln \left(1 - 2 r \cos(\theta_{T}) + r^2 \right).
\label{parte real de logaritmos}
\end{equation}
Thus, defining,
\begin{equation}
\begin{aligned}
    \Delta_{+} (p) &= 1 - 2 e^{-\beta (E_{\textbf{p}} + \mu)} \cos(\theta_{T}) + e^{-2 \beta (E_{\textbf{p}} + \mu) }, \\   \Delta_{-} (p) &= 1 - 2 e^{-\beta (E_{\textbf{p}} - \mu)} \cos(\theta_{T}) + e^{-2 \beta (E_{\textbf{p}} - \mu)},
\label{definicion de DELTAS}
\end{aligned}
\end{equation}
the real thermal component of (\ref{Z-H en terminos de logaritmos con densidad de energia de V}) can be expressed as,
\begin{equation}
    A_{\texttt{th}} = -\frac{V}{2} \int \frac{d^3 p}{(2 \pi)^3} \left[\ln \Delta_{+} (p) + \ln \Delta_{-} (p) \right].
\label{componente termica real 1}
\end{equation}
The corresponding phase can be written in terms of,
\begin{equation}
    \phi_{+} (p) = \arctan2 \left[e^{-\beta( E_{\textbf{p}} + \mu)} \sin (\theta_{T}) , 1 - e^{-\beta (E_{\textbf{p}} + \mu)} \cos (\theta_{T}) \right],
\label{componente real de Z-H 1}
\end{equation}
and,
\begin{equation}
      \phi_{-} (p) = \arctan2 \left[e^{-\beta( E_{\textbf{p}} - \mu)} \sin (\theta_{T}) , 1 - e^{-\beta (E_{\textbf{p}} - \mu)} \cos (\theta_{T}) \right].
\label{componente real de Z-H 1.1}
\end{equation}
Consequently, for the hybrid thermal component,
\begin{equation}
    B_{\texttt{th}} = V \int \frac{d^3 p}{(2 \pi)^3} \left[\phi_{+} (p) - \phi_{-} (p) \right].
\label{componente real de Z-H 2}
\end{equation}
Because \(r \pm < 1\) in the strict non-condensed domain,
\begin{equation}
    \text{Re} \, \left(1 - r_{\pm} e^{\pm i \theta_{T}} \right) > 0.
\label{argumentos delogaritmos permanecen en semiplano derecho}
\end{equation}
Therefore, these arguments also remain in the open right half-plane and the principal logarithm fixed above applies continuously. Accordingly, \(\phi_{\pm}\) in Eqs. (\ref{componente real de Z-H 1}) and (\ref{componente real de Z-H 1.1}) are uniquely defined by the corresponding principal arguments. Using the identity, 
\begin{equation}
    J^{+} (A + i B) + J^{-}(A-iB) = A + ij B, 
\end{equation}
the complete partition function acquires the Hermitian hypercomplex form,
\begin{equation}
    \ln Z_{\mathbb{H}} = A + ij B. 
\end{equation}
Thus, the hybrid \(ij\)-component has a precise spectral origin: it represents the relative thermal phase between the two conjugate idempotent sectors.

\subsection{Series representation of the thermal sector}

The physical content becomes particularly transparent by expanding the logarithms according to,
\begin{equation}
    -\ln (1-z) = \sum_{k=1}^{\infty} \frac{z^k}{k}, \hspace{1cm}  |z| < 1. 
\label{definicion de expansion de logaritmo}
\end{equation}
The \(J^{+}\) thermal contribution becomes,
\begin{equation}
    \ln Z_{{(+, \texttt{th})}} = 2 V \int \frac{d^3 p}{(2 \pi)^3} \sum_{k=1}^{\infty} \frac{e^{-k \beta E_{\textbf{p}}}}{k} \left[\cosh (k \beta \mu) \cos \left(\frac{k \beta \gamma}{2} \right) - i \sinh (k \beta \mu) \sin \left(\frac{k \beta \gamma}{2} \right)  \right],
\label{contribucion termica de Z+}
\end{equation}
and the \(J^{-}\) contribution is its complex conjugate,
\begin{equation}
    \ln Z_{{(-, \texttt{th})}} = 2 V \int \frac{d^3 p}{(2 \pi)^3} \sum_{k=1}^{\infty} \frac{e^{-k \beta E_{\textbf{p}}}}{k} \left[\cosh (k \beta \mu) \cos \left(\frac{k \beta \gamma}{2} \right) + i \sinh (k \beta \mu) \sin \left(\frac{k \beta \gamma}{2} \right)  \right].
\label{contribucion termica de Z-}
\end{equation}
Recombining both sectors (\ref{contribucion termica de Z+}) and (\ref{contribucion termica de Z-}) through the idempotent basis (\ref{bases idempotentes}) gives,
\begin{equation}
    \ln Z_{\mathbb{H}}^{\texttt{th}} = 2V \int \frac{d^3 p}{(2 \pi)^3} \sum_{k=1}^{\infty} \frac{e^{-k \beta E_{\textbf{p}}}}{k} \left[\cosh (k \beta \mu) \cos \left(\frac{k \beta \gamma}{2} \right) - i j \sinh (k \beta \mu) \sin \left(\frac{k \beta \gamma}{2} \right)  \right].
\label{Z-H en terminos de cosh y sinh}
\end{equation}
This expression exhibits two qualitatively different contributions. The real component is governed by \(\cosh (x) \cos (y) \), whereas the hybrid component is governed by \(-\sinh (x) \sin (y)\). Therefore, the \(ij\)-component requires simultaneously a nonzero dissipative deformation and a nonzero charge asymmetry. In particular,
\begin{equation}
    \begin{aligned}
        \gamma = 0  \hspace{0.5cm} \Rightarrow \hspace{0.5cm} B_{\texttt{th}} = 0, \\
        \mu= 0    \hspace{0.5cm}    \Rightarrow \hspace{0.5cm}  B_{\texttt{th}} = 0.
\label{limites gamma y mu = 0 parte 1}
    \end{aligned}
\end{equation}
The first limit corresponds to the disappearance of the hypercomplex thermal phase, whereas the second reflects particle-antiparticle symmetry. Specific cancellations may also occur for discrete values of \(\beta \gamma/2\); however, for generic \(\gamma \neq 0\) and \(\mu \neq 0\), one expects \(B_{\texttt{th}} \neq 0\). Thus, \(B_{\texttt{th}}\) encodes the joint dependence of the hybrid thermal sector on charge asymmetry and dissipation.

\vspace{0.2cm}
\noindent
Using the standard momentum integral \cite{tablas de integrales grad,tablas de integrales Abramowitz},
\begin{equation}
    \int \frac{d^3 p}{(2 \pi)^3} e^{- \beta k E_{\textbf{p}}} = \frac{m^2_{\texttt{eff}}}{2 \pi^2 k \beta} K_{2}(k \beta m_{\texttt{eff}}),
\label{integracion de e(k beta Ep)}
\end{equation}
where \(K_{2}(x)\) is the modified Bessel function of the second kind. Therefore, the thermal partition function becomes,
\begin{equation}
     \ln Z_{\mathbb{H}}^{\texttt{th}} = \frac{V m^2_{\texttt{eff}}}{\pi^2 \beta} \sum_{k=1}^{\infty} \frac{K_{2}(k \beta m_{\texttt{eff}})}{k^2} \left[\cosh (k \beta \mu) \cos \left(\frac{k \beta \gamma}{2} \right) - i j \sinh (k \beta \mu) \sin \left(\frac{k \beta \gamma}{2} \right)  \right].
\label{Z-H integrada}
\end{equation}
Eq. (\ref{Z-H integrada}) provides a compact representation of the generalized thermal sector. The dissipative parameter enters both through the effective mass and through an oscillatory thermal phase. These two effects have different analytic origins: \(m_{\texttt{eff}}\) modifies the quasiparticle spectrum, whereas the trigonometric factors arise from the imaginary displacement of the two conjugate chemical potentials.

\vspace{0.2cm}
\noindent
In the limit \(\gamma \rightarrow 0\), \(m_{\texttt{eff}} \rightarrow m\) and \(\mu_{\pm} \rightarrow \mu\), while the hybrid term vanishes continuously. Eq. (\ref{Z-H integrada}) therefore reduces to
\begin{equation}
    \ln Z_{\mathbb{H}}^{\texttt{th}} =  \frac{V m^2}{\pi^2 \beta} \sum_{k=1}^{\infty} \frac{K_{2}(k \beta m)}{k^2} \cosh (k \beta \mu), 
\label{estrucura termica estandar en U(1)}
\end{equation}
which is the standard thermal structure of a relativistic charged complex scalar field \cite{Kapusta,Alexander,Bellac,Landsman}. This provides a direct connection between the hypercomplex dissipative theory and equilibrium relativistic Bose statistics.

\subsection{Vacuum contribution and thermal finiteness}
The Matsubara summation separates the partition function naturally into a zero-point contribution and a finite-temperature contribution. The vacuum term is,
\begin{equation}
    \ln Z_{\texttt{vac}} = - \beta V \int \frac{d
^3 p}{(2 \pi)^3} E_{\textbf{p}},
\label{energia de punto cero 1}
\end{equation}
where \(E_{\textbf{p}}\) is defined in (\ref{energia modificada por la disipacion}). For large momenta we have,
\begin{equation}
    E_{\textbf{p}} = p + \frac{m^2_{\texttt{eff}}}{2p} + \mathcal{O}(p^{-3}),
\label{momento grande}
\end{equation}
then,
\begin{equation}
    \int \frac{d^3 p}{(2 \pi)^3} E_{\textbf{p}} \sim \int^{\Lambda} dp \, p^3.
\label{divergencia ultravioleta}
\end{equation}
Therefore, Eq. (\ref{divergencia ultravioleta}) contains the usual ultraviolet divergence associated with the zero-point energy of a relativistic quantum field. This divergence is independent of the thermal occupation factors and may be treated by the standard vacuum renormalization prescriptions employed in finite-temperature quantum field theory \cite{Kapusta,Landsman,regularizacion,regularizacion 2,renormalizacion energia cero,renormalizacion, renormalizacion 2, renormalizacion 3}. By contrast, every term in the thermal expansion contains \(e^{- k \beta E_{\textbf{p}}}\), and consequently behaves for \(p \rightarrow \infty\) as \(p^2 e^{- k \beta p}\); hence, \(\ln Z_{\mathbb{H}}^{\texttt{th}}\) is ultraviolet finite. The ultraviolet divergence of the complete Gaussian determinant is therefore confined to the vacuum contribution, while the dissipative thermal sector remains finite for nonzero temperature within the domain where the quasiparticle spectrum and the thermal expansion are well defined. After renormalizing the vacuum contribution, the complete renormalized partition function can therefore be written as,
\begin{small}
\begin{equation}
    \ln Z_{\mathbb{H}}^{\texttt{ren}} = \ln Z_{\texttt{Vac}}^{\texttt{ren}} +\ln Z_{0, \mathbb{H}} +   \frac{V m^2_{\texttt{eff}}}{\pi^2 \beta} \sum_{k=1}^{\infty} \frac{K_{2}(k \beta m_{\texttt{eff}})}{k^2} \left[\cosh (k \beta \mu) \cos \left(\frac{k \beta \gamma}{2} \right) - i j \sinh (k \beta \mu) \sin \left(\frac{k \beta \gamma}{2} \right) \right],
\label{Z-H renormalizada}
\end{equation}
\end{small}
where \(\ln Z_{\texttt{Vac}}^{\texttt{ren}}\) denotes the renormalized vacuum contribution, whereas \(\ln Z_{0, \mathbb{H}}\) denotes the contribution associated with the static zero-frequency and zero-momentum sector. Furthermore,
\begin{equation}
    \ln Z_{0, \mathbb{H}} = J^{+} \ln Z_{0,+} + J^{-} \ln Z_{0,-}.
\label{definicion de lnZ(0 H)}
\end{equation}
Choosing conjugate cycles \(Z_{0,-} = Z_{0,+}^{\ast} \) ensures that \(Z_{0, \mathbb{H}}\) also preserves the hypercomplex Hermitian structure. The expression (\ref{Z-H renormalizada}) summarizes the thermodynamic effect of the hypercomplex dissipative deformation. The equilibrium statistical contribution and the hypercomplex phase remain organized within a single Hermitian partition function, while the conventional relativistic Bose gas is continuously recovered when the dissipative parameter vanishes.

\subsection{Physical interpretation and outlook of the thermal phase}
The generalized partition function obtained here contains two independent dimensionless thermal structures: \( \beta \mu\) and \( \theta_{T}\). The first governs the conventional particle-antiparticle occupation asymmetry, while the second determines the relative phase between the projected \(J^{\pm}\) sectors. The generalized thermal factors may therefore be interpreted schematically as,
\begin{equation}
    e^{-\beta(E_{\textbf{p}} \mp \mu)} e^{\pm i\theta_{T}}.
\label{factores termales generalizados}
\end{equation}
The first factor represents ordinary Boltzmann suppression, whereas the second originates entirely from the hypercomplex dissipative structure. This separation is physically useful because it shows that the phase does not arise from an explicitly complex quasiparticle energy. The spectrum \(E_{\textbf{p}}\) (\ref{energia modificada por la disipacion}) remains real in the stable region, while complex phases appear only after projection of the complete hypercomplex theory onto its conjugate complex sectors. The underlying hypercomplex theory therefore remains Hermitian with respect to its native conjugation, even though the projected finite-temperature description exhibits structures that resemble those encountered in effective non-Hermitian systems.
Related phase structures occur in open and effectively non-Hermitian quantum systems, where coupling to environmental degrees of freedom, gain-loss mechanisms, projection procedures, or effective evolution operators can produce complex spectral or dynamical structures \cite{non hermitian fisica,super exponencial,nuevo tipo no hermitiano, hermitiano de non hermitiano, fases topologicas}. Non-Hermiticity has also been associated with modifications of thermal entanglement and information-thermodynamic behavior \cite{entrelazamiento termal,informacion termodinamica}. These connections are particularly suggestive in the present case because the complex thermal weights arise from the projection of an enlarged Hermitian algebra rather than from introducing non-Hermiticity as a fundamental assumption. Nevertheless, the appearance of the phase factor in Eq. (\ref{factores termales generalizados}) should not by itself be interpreted as evidence for a non-Hermitian thermodynamic phase. The present analysis establishes only the algebraic and thermal origin of the conjugate phase structure. Determining whether it can be associated with effective non-Hermitian phase transitions, exceptional spectral structures, thermal entanglement phenomena, or genuinely nonequilibrium behavior requires additional dynamical criteria that lie beyond the equilibrium partition-function construction considered here. On the other hand, the complementary static Fourier sector also remains an open problem. In particular, no identification of \(Z_{0}\) with a conventional Landau potential or Bose-Einstein condensate has been imposed in the present work. A detailed analysis of the stationary conditions and integration cycle of the static-sector hypercomplex action lies beyond the scope of the present work.

\section{Thermodynamic observables of \(Z_{\mathbb{H}}\)}
\label{seccion 5}
Having obtained the renormalized hypercomplex partition function, we now determine the corresponding thermodynamic observables. All thermodynamic derivatives are taken at fixed \(m\) and \(\gamma\). The thermal contribution derived in the previous section is,
\begin{equation}
    \ln Z_{\mathbb{H}}^{\texttt{th}} = \frac{V m^2_{\texttt{eff}}}{\pi^2 \beta} \sum_{k=1}^{\infty} \frac{K_{2}(x_{k})}{k^2} F_{k}; \hspace{1.5cm}  x_{k} = k \beta m_{\texttt{eff}}. 
\label{Z contribucion termica para observables}
\end{equation}
We have introduced the compact hypercomplex thermal factor,
\begin{equation}
    F_{k} = \cosh (k \beta \mu) \cos \left(\frac{k \beta \gamma}{2} \right) - i j \sinh (k \beta \mu) \sin \left(\frac{k \beta \gamma}{2} \right).
\label{definicion de F_{k}}
\end{equation}
The renormalized partition function may therefore be written as,
\begin{equation}
    \ln Z_{\mathbb{H}}^{\texttt{ren}} = \ln Z_{\texttt{Vac}}^{\texttt{ren}} +  \ln Z_{0,\mathbb{H}} + \ln Z_{\mathbb{H}}^{\texttt{th}}.
\label{funcion de particion completa renormalizada}
\end{equation}
The quantity \(\ln Z_{\texttt{Vac}}^{\texttt{ren}}\) denotes the renormalized vacuum contribution, whereas \(\ln Z_{0,\mathbb{H}}\) denotes the contribution from the static zero-frequency and zero-momentum sector. In the following we focus on bulk vacuum-subtracted thermodynamic densities. Within the strict non-condensed domain and in the thermodynamic limit, the isolated static Fourier mode contributes only \(O (V^{-1})\) corrections to these intensive quantities. Therefore,
\begin{equation}
    x_{\mathbb{H}}^{\texttt{sub}} = x_{\mathbb{H}}^{\texttt{th}} + O(V^{-1}), 
    \hspace{2cm}
    V \rightarrow \infty,
\label{observables termodimanicas esquematicamente}
\end{equation}
where \(x_{\mathbb{H}}\) denotes any bulk thermodynamic density. The static sector is nevertheless retained formally in the complete partition function because its behavior may become nontrivial near a macroscopically occupied or genuinely degenerate configuration.

\subsection{Thermodynamic potential and pressure}
The grand thermodynamic potential is defined in the standard way \cite{Kapusta,Alexander,Bellac},
\begin{equation}
    \Omega_{\mathbb{H}} = - \frac{1}{\beta} \ln Z_{\mathbb{H}}.
\label{potencial termodinamico en H}
\end{equation}
Consequently, the thermal contribution is,
\begin{equation}
    \Omega_{\mathbb{H}}^{\texttt{th}} = - \frac{V m^2_{\texttt{eff}}}{\pi^2 \beta^2} \sum_{k=1}^{\infty} \frac{K_{2}(x_{k})}{k^2} F_{k}.
\label{contribucion termica al potencial termodinamico en H}
\end{equation}
The generalized pressure follows from,
\begin{equation}
    P_{\mathbb{H}} = -\frac{ \Omega_{\mathbb{H}}}{V} = \frac{1}{\beta V} \ln Z_{\mathbb{H}},
\label{definicion de presion en H}
\end{equation}
and therefore, 
\begin{equation}
    P_{\mathbb{H}}^{\texttt{th}} = \frac{m^2_{\texttt{eff}}}{\pi^2 \beta^2} \sum_{k=1}^{\infty} \frac{K_{2}(x_{k})}{k^2} F_{k}.
\label{presion en H }
\end{equation}
Since the hypercomplex partition function is Hermitian-valued, the pressure can be separated as,
\begin{equation}
    P_{\mathbb{H}}^{\texttt{th}} = P_{\Re}^{\texttt{th}} + ij P_{ij}^{\texttt{th}}.
\label{presion separada en parte real y parte ij}
\end{equation}
In the equilibrium limit, \(P_{ij}^{\texttt{th}}=0\). The real component of (\ref{presion separada en parte real y parte ij}) is,
\begin{equation}
    P_{\Re}^{\texttt{th}} = \frac{m^2_{\texttt{eff}}}{\pi^2 \beta^2} \sum_{k=1}^{\infty} \frac{K_{2}( x_{k})}{k^2} \cosh (k \beta \mu) \cos \left(\frac{k \beta \gamma}{2} \right),
\label{parte real de la presion}
\end{equation}
and the complementary hybrid component is,
\begin{equation}
    P_{ij}^{\texttt{th}} = - \frac{m^2_{\texttt{eff}}}{\pi^2 \beta^2} \sum_{k=1}^{\infty} \frac{K_{2}(x_{k})}{k^2} \sinh (k \beta \mu) \sin \left(\frac{k \beta \gamma}{2} \right).
\label{parte hibrida de la presion}
\end{equation}
Thus, dissipation generates an oscillatory modulation of the thermal pressure while preserving the ordinary relativistic Bessel structure. The real and \(ij\)-components are not independent thermodynamic systems; they are generated by the recombination of the two conjugate \(J^{\pm}\) sectors of the same hypercomplex partition function.

\subsection{Conserved-charge density}
In the grand canonical ensemble,
\begin{equation}
    \frac{\partial \ln Z_{\mathbb{H}}}{\partial \mu} = \beta \left< Q\right>,
\label{definicion de numero de particulas}
\end{equation}
and for the conserved-charge density,
\begin{equation}
    n_{\mathbb{H}} = \frac{\left< Q\right>}{V}.
\label{numero de particulas en H}
\end{equation}
For convenience, introduce,
\begin{equation}
    G_{k} = \sinh (k \beta \mu) \cos \left(\frac{k \beta \gamma}{2}\right) - ij \cosh (k \beta \mu) \sin\left(\frac{k \beta \gamma}{2}\right).
\label{G_{k} para la presion}
\end{equation}
Then differentiation of Eq. (\ref{Z contribucion termica para observables}) gives,
\begin{equation}
    n_{\mathbb{H}}^{\texttt{th}} = \frac{m^2_{\texttt{eff}}}{\pi^2 \beta} \sum_{k=1}^{\infty} \frac{K_{2}(x_{k})}{k} G_{k}.
\label{numero de particulas en H derivada}
\end{equation}
Explicitly, the real part is,
\begin{equation}
    n_{\Re}^{\texttt{th}} = \frac{m^2_{\texttt{eff}}}{\pi^2 \beta} \sum_{k=1}^{\infty} \frac{K_{2}(x_{k})}{k} \sinh (k \beta \mu) \cos \left(\frac{k \beta \gamma}{2} \right),
\label{parte real de numero de particulas en H}
\end{equation}
and for the hybrid part, we obtain,
\begin{equation}
      n_{ij}^{\texttt{th}} = - \frac{m^2_{\texttt{eff}}}{\pi^2 \beta} \sum_{k=1}^{\infty} \frac{K_{2}(x_{k})}{k} \cosh (k \beta \mu) \sin \left(\frac{k \beta \gamma}{2} \right).
\label{parte hibrida de numero de particulas en H}
\end{equation}
The real conserved-charge density is odd under \(\mu \rightarrow -\mu\), as expected from particle-antiparticle symmetry. In particular,
\begin{equation}
    n_{\Re}^{\texttt{th}} (T, \mu=0, \gamma) = 0. 
\label{parte real de numero de particulas en H desvanecida}
\end{equation}
The hybrid component may remain nonzero at \(\mu=0\), since it characterizes the response of the full hypercomplex partition function rather than the ordinary real charge density. A detailed physical interpretation of this complementary response lies beyond the scope of the present work.

\subsection{Energy density}
The internal energy in the grand canonical ensemble follows from,
\begin{equation}
    U= - \partial_{\beta} \ln Z + \mu \left< Q\right>,
\label{definicion de energia interna}
\end{equation}
and therefore,
\begin{equation}
    \epsilon_{\mathbb{H}} = - \frac{1}{V} \partial_{\beta} \ln Z_{\mathbb{H}} + \mu n_{\mathbb{H}}.
\label{densidad de energia en H}
\end{equation}
Since both the Bessel functions and the generalized thermal factors depend explicitly on \(\beta\), the derivative must be taken before separating the real and hybrid sectors. We define, 
\begin{equation}
    H_{k} = \cosh (k \beta \mu) \sin \left(\frac{k \beta \gamma}{2} \right) + ij \sinh (k \beta \mu) \cos \left(\frac{k \beta \gamma}{2} \right).
\label{definicion de H_{k}}
\end{equation}
The derivative of \(F_{k}\) (\ref{definicion de F_{k}}) is,
\begin{equation}
    \frac{\partial F_{k}}{\partial \beta} = k \mu G_{k} - \frac{k \gamma}{2} H_{k}.
\label{derivada de F_{k}}
\end{equation}
Using the Bessel identity\footnote{ Derivative of the Bessel function \cite{tablas de integrales grad,tablas de integrales Abramowitz}: \(\frac{dK_{2}(x)}{dx} = - K_{1}(x) - \frac{2}{x} K_{2}(x)\).}, together with \(x_{k} = k \beta m_{\texttt{eff}}\), the terms proportional to \(\mu G_{k}\) cancel against the explicit \(\mu n\) contribution in Eq. (\ref{densidad de energia en H}). The resulting thermal energy density is,
\begin{equation}
    \epsilon_{\mathbb{H}}^{\texttt{th}} = \frac{m^2_{\texttt{eff}}}{\pi^2 \beta^2} \sum_{k=1}^{\infty} \frac{3 K_{2}(x_{k}) + x_{k} K_{1}(x_{k}) }{k^2} F_{k} + \frac{\gamma \,m^2_{\texttt{eff}}}{2 \pi^2 \beta} \sum_{k=1}^{\infty} \frac{K_{2}(x_{k})}{k} H_{k}.
\label{densidad de energia en H TOTAL}
\end{equation}
The first contribution retains the standard relativistic massive-boson structure, modified by the generalized factor \(F_{k}\). The second contribution originates exclusively from the explicit \(\beta\)-dependence of the dissipative thermal phase and therefore vanishes continuously when \(\gamma \rightarrow 0\). The real component of (\ref{densidad de energia en H TOTAL}) can be written as,
\begin{equation}
\begin{aligned}
    \epsilon_{\Re}^{\texttt{th}} &= \frac{m^2_{\texttt{eff}}}{\pi^2 \beta^2} \sum_{k=1}^{\infty} \frac{3 K_{2}(x_{k}) + x_{k} K_{1}(x_{k}) }{k^2}  \cosh (k \beta \mu) \cos \left(\frac{k \beta \gamma}{2} \right) \\
    &+ \frac{\gamma \,m^2_{\texttt{eff}}}{2 \pi^2 \beta} \sum_{k=1}^{\infty} \frac{K_{2}(x_{k})}{k} \cosh (k \beta \mu) \sin \left(\frac{k \beta \gamma}{2} \right)
\end{aligned}
\label{parte real de densitad de energia TOTAL}
\end{equation}
and for the hybrid part as,
\begin{equation}
\begin{aligned}
    \epsilon_{ij}^{\texttt{th}}& = - \frac{m^2_{\texttt{eff}}}{\pi^2 \beta^2} \sum_{k=1}^{\infty} \frac{3 K_{2}(x_{k}) + x_{k} K_{1}(x_{k}) }{k^2}  \sinh (k \beta \mu) \sin \left(\frac{k \beta \gamma}{2} \right)\\
   & + \frac{\gamma \,m^2_{\texttt{eff}}}{2 \pi^2 \beta} \sum_{k=1}^{\infty} \frac{K_{2}(x_{k})}{k} \sinh (k \beta \mu) \cos \left(\frac{k \beta \gamma}{2} \right).
\end{aligned}
\label{parte hibrida de densitad de energia TOTAL}
\end{equation}
The presence of the second term in Eqs. (\ref{densidad de energia en H TOTAL})-(\ref{parte hibrida de densitad de energia TOTAL}) shows that the dissipative thermal phase affects the internal energy not only through the modified occupation factors, but also through its explicit temperature dependence.

\subsection{Entropy and thermodynamic consistency}
\label{5.4 entropia y consistencia termodinamica}
The entropy density can be obtained from,
\begin{equation}
    s_{\mathbb{H}} = \beta \left(\epsilon_{\mathbb{H}} + P_{\mathbb{H}} - \mu n_{\mathbb{H}} \right).
\label{definicion de entropia en H}
\end{equation}
For the thermal contribution one finds,
\begin{equation}
\begin{aligned}
    s_{\mathbb{H}}^{\texttt{th}} = \frac{m^2_{\texttt{eff}}}{\pi^2 \beta} \sum_{k=1}^{\infty} \frac{4 K_{2}(x_{k}) + x_{k} K_{1}(x_{k})}{k^2} F_{k} - \frac{\mu \, m^2_{\texttt{eff}}}{\pi^2} \sum_{k=1}^{\infty} \frac{K_{2}(x_{k})}{k} G_{k} + \frac{\gamma \,m^2_{\texttt{eff}}}{2 \pi^2} \sum_{k=1}^{\infty} \frac{K_{2}(x_{k})}{k} H_{k}.
\label{contribucion termal de entropia en H}
\end{aligned}
\end{equation}
The usual Euler relation for homogeneous thermodynamics follows identically,
\begin{equation}
    \epsilon_{\mathbb{H}} + P_{\mathbb{H}} = T s_{{\mathbb{H}}} + \mu n_{\mathbb{H}}.
\label{relacion de Gibbs Duhem}
\end{equation}
Thus, the hypercomplex extension does not require a modification of the fundamental grand-canonical thermodynamic identities. The generalized structure is carried by the values of the observables themselves rather than by a deformation of the Legendre relations.

\subsection{Physical domain of the thermal sector}
The thermal construction requires a real quasiparticle spectrum \(E_{\textbf{p}}\), which in turn requires \(m_{\texttt{eff}}^{2} =( m^2 - \frac{\gamma^2}{4}) > 0\) or equivalently,
\begin{equation}
    |\gamma| < 2m.
\label{condicion para region dinamica estable}
\end{equation}
This condition defines the dynamically stable region in which all thermal quasiparticle energies remain real. The series representation of the logarithms additionally requires, 
\begin{equation*}
     \begin{vmatrix}
e^{-\beta (E_{\textbf{p}} - \mu_{\mp})}\end{vmatrix} < 1. 
\end{equation*}
Since the imaginary dissipative part of \(\mu_{\pm}\) contributes only a phase,
\begin{equation}
    \begin{vmatrix}
e^{-\beta (E_{\textbf{p}} - \mu_{\mp})}\end{vmatrix} = e^{-\beta(E_{\textbf{p}} - \mu)}.
\end{equation}
The non-condensed thermal sector is therefore well defined for
\begin{equation}
    |\mu| < m_{\texttt{eff}}.
\label{condicion mu < m_eff}
\end{equation}
Within the strict non-condensed domain (\ref{condicion mu < m_eff}), all eigenvalues remain nonzero. At the boundary \(|\mu| = m_{\texttt{eff}}\), the \(\mathbf{p} = 0\) sector must be treated separately. A true zero eigenvalue may occur only at the additional discrete resonance condition \(\omega_{n} = s \, \gamma / 2\). Away from these resonant values, the loss of absolute convergence of the thermal series at the boundary does not by itself imply a zero eigenvalue. A complete treatment of this boundary problem, including possible zero-eigenvalue configurations and the construction of a hypercomplex effective potential, lies beyond the scope of the present analysis.

\section{Recovery of the relativistic Bose gas}
\label{seccion 6}
A fundamental consistency requirement of the formalism is the recovery of ordinary relativistic finite-temperature field theory when the dissipative deformation is removed. Taking the limit \(\gamma \rightarrow 0\), one obtains, 
\begin{equation*}
\begin{matrix}
m_{\texttt{eff}} \rightarrow  m, \hspace{0.5cm} & \hspace{0.5cm} \mu_{\pm} \rightarrow \mu,  \\
\cos \left(\frac{k \beta \gamma}{2}\right)\rightarrow 1, \hspace{0.5cm} & \hspace{0.5cm} \sin \left(\frac{k \beta \gamma}{2}\right) \rightarrow 0.  
\end{matrix}
\end{equation*}
Consequently, \(F_{k} \rightarrow \cosh (k \beta \mu)\), and all \(ij\)-components vanish. The thermal partition function becomes,
\begin{equation}
    \ln Z_{\texttt{Bose}}^{\texttt{th}} = \frac{V m^2}{\pi^2 \beta} \sum_{k=1}^{\infty} \frac{K_{2}(k \beta m)}{k^2} \cosh(k \beta \mu), 
\label{funcion de particion Bose parte thermal}
\end{equation}
which is the conventional result for a relativistic charged complex scalar gas \cite{Kapusta,Alexander,Bellac,Landsman}. The corresponding pressure is,
\begin{equation}
    P_{\texttt{Bose}}^{\texttt{th}} = \frac{m^2}{\pi^2 \beta^2} \sum_{k=1}^{\infty} \frac{K_{2}(k \beta m)}{k^2} \cosh(k \beta \mu),
\label{presion de Bose}
\end{equation}
the charge density is,
\begin{equation}
    n_{\texttt{Bose}}^{\texttt{th}} = \frac{m^2}{\pi^2 \beta} \sum_{k=1}^{\infty} \frac{K_{2}(k \beta m)}{k} \sinh(k \beta \mu),
\label{densidad de carga de Bose}
\end{equation}
and the energy density becomes,
\begin{equation}
     \epsilon_{\texttt{Bose}}^{\texttt{th}} = \frac{m^2}{\pi^2 \beta^2} \sum_{k=1}^{\infty} \frac{3K_{2}(k \beta m) + k \beta m K_{1}(k \beta m)}{k^2} \cosh(k \beta \mu).
\label{densidad de energia de Bose}
\end{equation}
Therefore, the hypercomplex construction connects continuously with the standard equilibrium theory when dissipation is removed. This recovery applies simultaneously to the partition function, pressure, charge density, energy density, entropy, and equation of state, and constitutes a non-trivial consistency condition of the complete construction. Another useful limit is \(\mu \rightarrow 0\). In this case, \(\sinh(k \beta \mu) \rightarrow 0\), and therefore,
\begin{equation}
    P_{ij}^{\texttt{th}} \rightarrow 0, \hspace{1cm}  \epsilon_{ij}^{\texttt{th}} \rightarrow 0,  \hspace{1cm} n_{\Re}^{\texttt{th}} \rightarrow 0, 
    \hspace{1cm}
    s_{ij}^{\texttt{th}} \rightarrow 0.
\end{equation}
Thus, the thermodynamic potential, pressure, energy density, and entropy density become real in the charge-symmetric sector even for finite \(\gamma\). The disappearance of the \(ij\)-component at \(\mu = 0\) reflects the fact that the thermal phase contributing to these observables requires the simultaneous presence of charge asymmetry and dissipative deformation.

\section{Thermal regimes}
\label{seccion 7}
\subsection{High-temperature regime}
At sufficiently high temperatures,
\begin{equation}
    \beta m_{\texttt{eff}} \ll 1,
    \hspace{1cm}
    \beta |\mu| \ll 1,
    \hspace{1cm}
    \beta |\gamma| \ll 1. 
\label{regimen de alta temperatura}
\end{equation}
The modified Bessel function admits the asymptotic expansion\footnote{Expansion of the Bessel function \cite{tablas de integrales grad,tablas de integrales Abramowitz}: \(K_{2}(x) = \frac{2}{x^2} - \frac{1}{2} + \mathcal{O}(x^2 \ln x)\).}. In this regime, the leading real contribution to the pressure is\footnote{\(P_{\Re}^{\texttt{th}} \simeq \frac{2}{\pi^2 \beta^4} \sum_{k=1}^{\infty} \frac{1}{k^4}\), using: \(\sum_{k=1}^{\infty} \frac{1}{k^4} = \frac{\pi^4}{90}\).}
\begin{equation}
     P_{\Re}^{\texttt{th}} \quad \rightarrow \quad \frac{\pi^2}{45} T^4.
\label{presion a Alta temperatura}
\end{equation}
The energy density becomes,
\begin{equation}
    \epsilon_{\Re}^{\texttt{th}} \rightarrow \frac{\pi^2}{15} T^4 \hspace{0.6cm}  \Rightarrow  \hspace{0.6cm}   \epsilon_{\Re}^{\texttt{th}} = 3  P_{\Re}^{\texttt{th}}.
\label{densidad de energia a Alta temperatura}
\end{equation}
Similarly,
\begin{equation}
     s_{\Re}^{\texttt{th}} \quad \rightarrow \quad \frac{4\pi^2}{45} T^3.
\label{entropia a Alta temperatura}
\end{equation}
These are the Stefan-Boltzmann expressions corresponding to one complex scalar field, or equivalently two real bosonic degrees of freedom \cite{Kapusta,Alexander,Bellac,Landsman,piers}. Thus, at sufficiently high temperature, the dominant contribution is the conventional \(T^4\) relativistic behavior. Dissipative corrections become subleading in powers of the dimensionless quantities \(m_{\texttt{eff}}/T\), \(\mu/T\), and \(\gamma/T\).

\subsection{Low-temperature regime}
For
\begin{equation}
    \beta \left( m_{\texttt{eff}} - |\mu| \right) \gg 1;
\label{regimen de baja temperatura}
\end{equation}
the thermal pressure therefore behaves as\footnote{In this regime the modified Bessel function behaves as \cite{tablas de integrales grad,tablas de integrales Abramowitz}: \(K_{2}(x) \sim \sqrt{\frac{\pi}{2x}} e^{-x} \left(1 + \frac{15}{8x} + \cdots \right) \); the dominant contribution arises from \(k=1\).},
\begin{equation}
    P_{\mathbb{H}}^{\texttt{th}} \simeq 2T \left(\frac{ T \,m_{\texttt{eff}}}{2 \pi}\right)^{3/2} e^{-m_{\texttt{eff}}/T} \left[\cosh \left(\frac{\mu}{T}\right) \cos \left(\frac{\gamma}{2T}\right) - ij \sinh \left(\frac{\mu}{T}\right)  \sin \left(\frac{\gamma}{2T}\right) \right].
\label{presion a Baja temperatura}
\end{equation}
Combining the Boltzmann factor with the hyperbolic functions, the thermal envelope is bounded by \(e^{-( m_{\texttt{eff}} - |\mu|)/T}\), which vanishes for \(|\mu| < m_{\texttt{eff}}\), while the hypercomplex deformation generates an additional oscillatory phase controlled by \(\gamma/2T\). Hence, the exponential envelope dominates the low-temperature limit, and
\begin{equation}
    P_{\mathbb{H}}^{\texttt{th}}, \; n_{\mathbb{H}}^{\texttt{th}}, \; s_{\mathbb{H}}^{\texttt{th}}, \;\epsilon_{\mathbb{H}}^{\texttt{th}} \rightarrow 0 \qquad (T \rightarrow 0). 
\label{P_H, n_H, epsilon_H, s_H tienden a cero en Baja temperatura}
\end{equation}
Although the trigonometric phase oscillates increasingly rapidly as \(T \rightarrow 0\), its amplitude is exponentially suppressed. Therefore, no additional thermal ultraviolet or low-temperature divergence is generated by this phase.

\section{Weak dissipative deformation}
\label{seccion 8}
The hypercomplex thermal structure allows a controlled expansion around the conventional Bose gas. For weak dissipation,
\begin{equation*}
    |\gamma| \ll m,
\end{equation*}
the effective mass satisfies,
\begin{equation}
    m_{\texttt{eff}} = m - \frac{\gamma^2}{8m} + \mathcal{O}(\gamma^4).
\label{masa efectiva con pequena disipacion}
\end{equation}
Therefore, the quasiparticle spectrum receives no linear correction in \(\gamma\). For \(E_{\textbf{p}}^{(0)} = \sqrt{\textbf{p}^2 + m^2}\), one obtains,
\begin{equation}
    E_{\textbf{p}} = E_{\textbf{p}}^{(0)} - \frac{\gamma^2}{8 E_{\textbf{p}}^{(0)}} + \mathcal{O} (\gamma^4).
\label{energia con pequena disipacion}
\end{equation}
The thermal phase, on the other hand, contains,
\begin{equation}
    \cos \left( \frac{k \beta \gamma}{2} \right) = 1 - \frac{k^2 \beta^2 \gamma^2}{8} + \mathcal{O} (\gamma^4), \hspace{1.5cm} \sin \left( \frac{k \beta \gamma}{2} \right) = \frac{k \beta \gamma}{2} + \mathcal{O} (\gamma^3).
\label{cos y sin de k beta gamma con pequena disipacion}
\end{equation}
It follows immediately that the real thermal sector satisfies the perturbative structure,
\begin{equation}
    X_{\Re} (T, \gamma, \mu) = X_{\texttt{eq}} (T, \mu) + \gamma^2 X_{\Re}^{(2)} (T, \mu) + \mathcal{O} (\gamma^4),
\label{estructura perturbativa real con gamma pequeno}
\end{equation}
where \(X_{\texttt{eq}}\) denotes the conventional equilibrium result, while \(X_{\Re}^{(2)}\) and \(X_{ij}^{(1)}\) denote the leading dissipative corrections in the real and hybrid sectors, respectively. The complementary hypercomplex sector behaves as,
\begin{equation}
    X_{ij} (T, \gamma, \mu) = \gamma X_{ij}^{(1)} (T, \mu) + \mathcal{O} (\gamma^3).
\label{estructura perturbativa hibrida con gamma pequeno}
\end{equation}
Thus, the leading dissipative effect is not the same in the two components of the hypercomplex thermodynamic function. The hybrid thermal phase responds linearly to the dissipative parameter, whereas the ordinary real thermodynamic sector is perturbatively protected from linear corrections and begins at second order. This hierarchy has a simple origin. The spectral deformation
\(m^2 \rightarrow (m^2 -\frac{\gamma^2}{4})\) is even under \(\gamma \rightarrow - \gamma\), whereas the relative phase of the conjugate idempotent sectors changes sign,
\begin{equation}
    X_{\Re}(\gamma) = X_{\Re}(-\gamma) , \hspace{1.5cm}    X_{ij}(\gamma) = - X_{ij}(-\gamma), 
\end{equation}
within the thermal sector. This property provides a useful characterization of the hypercomplex dissipative deformation: weak dissipation first appears as a relative phase between conjugate thermal sectors, while modifications of ordinary real thermodynamic quantities arise only at the next perturbative order.

\section{Massless and gapless limits}
\label{seccion 9}
The massless limit requires some care because the stability scale determined by \(m^2_{\texttt{eff}} = m^2 - \frac{\gamma^2}{4}\). The limit \(m \rightarrow 0\) at fixed nonzero \(\gamma\) would imply \(m^2_{\texttt{eff}} < 0\), and therefore lies outside the stable quasiparticle region considered here. Within the non-condensed thermal domain, the formal limit \(m_{\texttt{eff}} \rightarrow 0\) also requires \(\mu \rightarrow 0\). The projected oscillatory factors associated with finite \(\gamma\) may nevertheless remain nontrivial. The conventional massless relativistic gas is consequently recovered through the simultaneous equilibrium limit,
\begin{equation*}
    m \rightarrow 0 , \hspace{1cm}
    \gamma \rightarrow 0 , \hspace{1cm}
    \mu \rightarrow 0.
\end{equation*}
Under these conditions,
\begin{equation}
    P = \frac{\pi^2}{45} T^4  , \hspace{0.5cm}
    \epsilon = \frac{\pi^2}{15} T^4  , \hspace{0.5cm}
    s = \frac{4 \pi^2}{45} T^3  , \hspace{0.5cm}
    n=0.
\end{equation}
This should be distinguished from the formal gapless condition,
\begin{equation*}
    m_{\texttt{eff}} \quad \rightarrow \quad 0
\end{equation*}
at finite \(\gamma\), which may retain a nontrivial thermal phase and therefore does not coincide with the ordinary massless Bose gas.

\vspace{0.2cm}
\noindent

\newpage
\section{Conclusions}
\label{conclusiones}
In this work, we constructed the finite-temperature partition function of a dissipative hypercomplex field theory, using the imaginary-time path-integral formalism. The idempotent decomposition leads to two conjugate complex sectors characterized by a real quasiparticle spectrum \(E_{\textbf{p}}\) with effective mass \(m_{\texttt{eff}}\), and by conjugate thermal parameters \(\mu_{\pm}\). The resulting Gaussian functional integral provides a consistent determinant representation and yields closed expressions for the pressure, charge density, energy density, and entropy. The conventional relativistic Bose gas is recovered continuously in the limit \(\gamma \rightarrow 0\), while finite dissipation generates an additional conjugate thermal phase and a characteristic perturbative hierarchy: the complementary \(ij\)-sector appears at first order in \(\gamma\), whereas corrections to the real thermodynamic sector begin at second order. These results establish the hypercomplex ring as a consistent algebraic framework for extending finite-temperature bosonic field theory while preserving its equilibrium limit. 

\vspace{0.2cm}
\noindent
Future work will address the static Fourier sector and its possible phase structure, fermionic extensions of the hypercomplex thermal formalism, generalized thermalization mechanisms, ergodicity in the \(\mathbb{H}\)-ring, and potential applications in technological and engineering contexts. A central direction will be the development of a more general and mathematically rigorous formulation of partition functions and path integrals within the \(\mathbb{H}\)-ring, including the characterization of functional measures, integration cycles, convergence conditions, Gaussian and non-Gaussian functional integrals, and their relation to the underlying hypercomplex algebra. Particular attention will also be given to the connection between the projected conjugate sectors and effective descriptions of open, non-equilibrium, and non-Hermitian quantum systems. In this context, it will be important to determine whether the thermal phase emerging from the \(\mathbb{H}\)-ring can provide an algebraic route to effective non-Hermitian phenomena while preserving Hermiticity at the level of the complete hypercomplex theory. Such investigations may further help clarify possible links with non-Hermitian thermal entanglement, information thermodynamics, and finite-temperature nonequilibrium quantum field theory.


\newpage
\begin{appendices}
\section{Real representation of the complex Gaussian functional integrals}
\label{apendice a}

\subsection*{Purpose of the appendix}
The functional integrals appearing in the idempotent decomposition of the hypercomplex partition function are naturally written in terms of complex fields. Although this representation is convenient for exploiting the algebraic structure of the hypercomplex ring and for performing the Matsubara decomposition, the underlying functional measure may equivalently be expressed in terms of real field components. The purpose of this appendix is to establish this correspondence explicitly. Writing a functional integral as a Gaussian integral over real variables does not by itself imply that the corresponding real quadratic form is positive definite. Positive definiteness instead concerns the convergence of the chosen integration cycle. The real-component representation establishes the Gaussian character of the projected functional integrals, whereas convergence requires an appropriate choice of complex integration cycle.
\vspace{0.2cm}
\noindent
The hypercomplex field is decomposed as,
\begin{equation}
    \Omega = J^{+} \Omega^{+} + J^{-} \Omega^{-};
\label{A1}
\end{equation}
where \(\Omega^{\pm}\) are ordinary complex fields. This is the same idempotent decomposition introduced in the main text. Under hypercomplex conjugation,
\begin{equation}
    \overline{\Omega} = J^{-} \overline{\Omega^{+}} + J^{+} \overline{\Omega^{-}}.
\label{A2}
\end{equation}
Since the present appendix is concerned with the determinant dependence and the convergence structure of the projected Gaussian, overall numerical factors that are independent of the thermodynamic parameters are absorbed into the normalization of the functional measure. Consequently, the two projected quadratic actions have the generic form,
\begin{equation}
    S_{+} = \int_{x} \overline{\Omega^{-}} (x) \mathcal{D}_{+} \Omega^{+}(x),
\label{A3}
\end{equation}
and,
\begin{equation}
     S_{-} = \int_{x} \overline{\Omega^{+}} (x) \mathcal{D}_{-} \Omega^{-}(x),
\label{A4}
\end{equation}
where
\begin{equation*}
    \int_{x} \equiv \int_{0}^{\beta} d \tau \int d^3 x. 
\end{equation*}
For the Euclidean operator derived in the main text,
\begin{equation}
    \mathcal{D}_{\pm} = - \left(\partial_{\tau} + \mu_{\pm} \right)^2 - \nabla^2 + m^2_{\texttt{eff}},
\label{A5}
\end{equation}
with,
\begin{equation}
    \mu_{\pm} = \mu \mp i \frac{\gamma}{2}, \hspace{1.5cm}   m^2_{\texttt{eff}} = m^2 - \frac{\gamma^2}{4}.
\label{A6}
\end{equation}
For real \(\mu\) and \(\gamma\),
\begin{equation*}
    \mu_{-} = \mu_{+}^{\ast}.
\end{equation*}

\subsection{Decomposition into real fields}
Every complex field may be written uniquely in terms of two real fields. We define,
\begin{equation}
    \Omega^{+} = \frac{1}{\sqrt{2}} (\phi_1 + i \phi_2),
\label{A7}
\end{equation}
and,
\begin{equation}
      \Omega^{-} = \frac{1}{\sqrt{2}} (\chi_1 + i \chi_2),
\label{A8}
\end{equation}
where \((\phi_{1,2}, \chi_{1,2}) \in \mathbb{R}\). Therefore,
\begin{equation*}
    \overline{\Omega^{+}} = \frac{1}{\sqrt{2}} (\phi_1 - i \phi_2),
\end{equation*}
and,
\begin{equation*}
      \overline{\Omega^{-}} = \frac{1}{\sqrt{2}} (\chi_1 - i \chi_2).
\end{equation*}
The functional measure transforms as,
\begin{equation}
    \mathcal{D} \Omega^{+} \mathcal{D} \overline{\Omega^{+}} \mathcal{D} \Omega^{-} \mathcal{D} \overline{\Omega^{-}} = \mathcal{N} \prod_{x} d \phi_1 d \phi_2 d \chi_1 d \chi_2,
\label{A9}
\end{equation}
where \(\mathcal{N}\) is a field-independent Jacobian. Since the transformation is linear, this Jacobian is a constant and can be absorbed into the overall normalization of \(Z_{\mathbb{H}}\). Thus, at the level of the functional measure, the two complex fields are equivalently represented by four real field components.

\subsection{Real representation of the \(J^+\) sector}
Consider
\begin{equation*}
    S_{+} = \int_{x} \overline{\Omega^{-}} \mathcal{D}_{+} \Omega^{+}.
\end{equation*}
Substituting (\ref{A7})-(\ref{A8}),
\begin{equation}
    S_{+}= \frac{1}{2} \int_{x} (\chi_1 - i \chi_2) \mathcal{D}_{+} (\phi_1 + i \phi_2).
\label{A11}
\end{equation}
Write the complex differential operator as,
\begin{equation}
    \mathcal{D}_{+} = A+ iB,
\label{A12}
\end{equation}
where \(A\) and \(B\) are real differential operators. From,
\begin{equation}
    \mathcal{D}_{+} = - \left(\partial_{\tau} + \mu - i \frac{\gamma}{2} \right)^2 - \nabla^2 + m^2_{\texttt{eff}},
\label{A13}
\end{equation} 
one obtains explicitly,
\begin{equation}
    A = - \left( \partial_{\tau} + \mu \right)^2 - \nabla^2 + m^2_{\texttt{eff}} + \frac{\gamma^2}{4},
\label{A13}
\end{equation}
and,
\begin{equation}
    B = \gamma \left(\partial_{\tau} + \mu \right).
\label{A14}
\end{equation}
Using
\begin{equation*}
    m^2_{\texttt{eff}} + \frac{\gamma^2}{4} = m^2,
\end{equation*}
Eq. (\ref{A13}) can also be written as,
\begin{equation}
     A = - \left( \partial_{\tau} + \mu \right)^2 - \nabla^2 + m^2.
\label{A15}
\end{equation}
Thus (\ref{A12}) acting on \(\Omega^{+}\),
\begin{equation}
    (A + iB) (\phi_1 + i \phi_2) = (A \phi_1 - B \phi_2) + i (B \phi_1 + A \phi_2).
\label{A16}
\end{equation}
Multiplication by \(\chi_1 - i \chi_2\) gives,
\begin{equation}
    S_{+} = \frac{1}{2} \int_{x} \left\{\chi_{1} (A \phi_1 - B \phi_2) + \chi_2 (B \phi_1 + A \phi_2) + i \left[\chi_1 (B \phi_1 +A \phi_2 ) - \chi_2 (A \phi_1 - B \phi_2)  \right] \right\}.
\label{A17}
\end{equation}
Eq. (\ref{A17}) therefore involves only the four real fields \(\phi_{1,2}\) and \(\chi_{1,2}\).

\subsection{Matrix form}
Introduce the real two-component vectors,
\begin{equation}
    \Phi = \begin{pmatrix}
\phi_1 \\ \phi_2
\end{pmatrix}, \hspace{1cm}
X= \begin{pmatrix}
\chi_1 \\ \chi_2
\end{pmatrix}.
\label{A18}
\end{equation}
Define the realification of \(\mathcal{D}_{+}\) by,
\begin{equation}
    \Re(\mathcal{D}_{+})= \begin{pmatrix}
A & -B  \\
B & A \\
\end{pmatrix}.
\label{A19}
\end{equation}
Then,
\begin{equation}
    S_{+} = \frac{1}{2} \int_{x} X^{T} \left[\Re(\mathcal{D}_{+}) + i J \Re(\mathcal{D}_{+}) \right] \Phi,
\label{A20}
\end{equation}
where
\begin{equation*}
  J=  \begin{pmatrix}
0 & 1  \\
-1 & 0 \\
\end{pmatrix}.
\end{equation*}
An equivalent and more useful representation follows by combining all real fields into,
\begin{equation}
    \Xi = \begin{pmatrix}
\phi_1 \\
\phi_2 \\
\chi_1 \\
\chi_2
\end{pmatrix}.
\label{A21}
\end{equation}
Then,
\begin{equation}
    S_{+} = \frac{1}{2} \int_{x} \Xi^{T} \mathbb{K}_{+} \Xi \;,
\label{A22}
\end{equation}
with the complex symmetric \(4 \times 4\) operator,
\begin{equation}
\mathbb{K}_{+} = \frac{1}{2}  \begin{pmatrix}
0 & 0 & A^T + iB^T & B^T-iAT \\
0 & 0 & -B^T + iA^T & A^T +iB^T \\
A+iB & -B+iA & 0  & 0  \\
B-iA & A+iB & 0 & 0 \\
\end{pmatrix},
\label{A23}
\end{equation}
up to integration-by-parts transposition of the differential entries. The precise symmetric representative may equivalently be written abstractly as,
\begin{equation}
    \mathbb{K}_{+} = \frac{1}{2} \begin{pmatrix}
0 & \textup{Q}_{+}^{T} \\
\textup{Q}_{+} & 0  \\
\end{pmatrix},
\label{A24}
\end{equation}
where \(\textup{Q}_{+} \equiv \Re(\mathcal{D}_{+}) + i J \Re(\mathcal{D}_{+})\) with the upper-right block understood as the formal transpose of \(\textup{Q}_{+}\) under the periodic boundary conditions. The real-component matrix \( \mathbb{K}_{+}\) is introduced only to display explicitly the quadratic character of the functional integral. It should not be identified with the reduced complex Hessian whose determinant defines \(\Det \mathcal{D}_+\). For the modal determinant and steepest-descent analysis we return to the independent bilinear complex coordinates, for which \(S_{\alpha} = \lambda_{\alpha} x_{\alpha} y_{\alpha} \). Consequently,
\begin{equation}
    Z_{+} = \mathcal{N}_{+} \int
 \mathcal{D} \phi_1  \mathcal{D} \phi_2  \mathcal{D} \chi_1  \mathcal{D} \chi_2 \exp \left[-\frac{1}{2} \int_{x} \Xi^{T} \mathbb{K}_{+} \Xi \right].
\label{A25}
\end{equation}
Eq. (\ref{A25}) establishes the Gaussian character of the projected functional integral over real fields. This does not imply that \(\mathbb{K}_{+}\) is a real positive-definite matrix: although the integration variables \((\phi_{1,2}, \chi_{1,2})\) are real, \(\mathbb{K}_{+}\) is generally complex because  \(\mathcal{D}_{+}\) contains the imaginary contribution generated by dissipation. Gaussian character and positive definiteness must therefore be treated separately.

\subsection{The \(J^{-}\) sector}
Because \(\mu_{-} = \mu_{+}^{\ast}\), we have, with the appropriate transpose under periodic boundary conditions,
\begin{equation}
    \mathcal{D}_{-} = \mathcal{D}_{+}^{\ast}.
\label{A27}
\end{equation}
The corresponding real quadratic operator therefore satisfies,
\begin{equation}
    \mathbb{K}_{-} =  \mathbb{K}_{+}^{\ast}.
\label{A28}
\end{equation}
Thus,
\begin{equation}
    Z_{-} = Z_{+}^{\ast},
\label{A29}
\end{equation}
provided that the integration cycles are chosen as complex-conjugate cycles. Consequently,
\begin{equation*}
    Z_{\mathbb{H}} = J^{+} Z_{+} + J^{-} Z_{-},
\end{equation*}
takes the form,
\begin{equation}
     Z_{\mathbb{H}} = J^{+} Z_{+} + J^{-} Z_{+}^{\ast}.
\label{A30}
\end{equation}
Writing,
\begin{equation*}
    Z_{+} = Re^{i \Theta},
\end{equation*}
one immediately obtains,
\begin{equation*}
    Z_{\mathbb{H}} = R \left(J^{+} e^{i \Theta} + J^{-} e^{-i \Theta} \right),
\end{equation*}
and therefore,
\begin{equation}
    Z_{\mathbb{H}} = R \left[\cos \Theta + ij \sin \Theta \right].
\label{A31}
\end{equation}
Hence the Hermitian hypercomplex structure follows directly from the conjugate relation between the two real Gaussian representations.

\subsection{Fourier-mode representation}
The previous argument becomes particularly transparent after Fourier expansion. For bosonic Matsubara modes,
\begin{equation}
    \Omega^{\pm} (\tau, \textbf{x}) = \sqrt{\frac{\beta}{V}} \sum_{n, \textbf{p}} e^{i(\omega_{n}\tau + \textbf{p} \cdot \textbf{x} )} \Omega^{\pm}_{n, \textbf{p}}.
\label{A32}
\end{equation}
For each mode,
\begin{equation*}
    \mathcal{D}_{\pm} \rightarrow \lambda_{\pm} (n, \textbf{p}), 
\end{equation*}
where
\begin{equation}
    \lambda_{\pm} = (\omega_{n} - i \mu_{\pm} )^2 + E^{2}_{\textbf{p}}.
\label{A33}
\end{equation}
Write,
\begin{equation*}
    \lambda_{+} = a_{n \textbf{p}} + i b_{n \textbf{p}}.
\end{equation*}
Using
\begin{equation}
    \mu_{+} = \mu - i \frac{\gamma}{2},
\end{equation}
one obtains,
\begin{equation}
    a_{n \textbf{p}} = \left(\omega_{n} - \frac{\gamma}{2}\right)^2 - \mu^2 + E^2_{\textbf{p}},
\label{A34}
\end{equation}
and,
\begin{equation}
    b_{n \textbf{p}} = -2 \mu \left(\omega_{n} - \frac{\gamma}{2}\right).
\label{A35}
\end{equation}
Each complex multiplication,
\begin{equation*}
    z \mapsto  \lambda_{+} z,
\end{equation*}
is represented on \(\mathbb{R}^2\) by,
\begin{equation}
    R(\lambda_{+}) = \begin{pmatrix}
a_{n\textbf{p}} & -b_{n\textbf{p}} \\
b_{n\textbf{p}} & a_{n\textbf{p}} \\
\end{pmatrix}.
\label{A36}
\end{equation}
Its determinant is,
\begin{equation}
    \det R(\lambda_{+}) = a_{n\textbf{p}}^2 + b_{n\textbf{p}}^2 = |\lambda_{+}|^2.
\label{A37}
\end{equation}
This is the standard realification identity. At finite regulator, the corresponding product of modal determinants satisfies,
\begin{equation}
 \det \Re(\mathcal{D}_{+}) = |\det \mathcal{D}_{+}|^2.
\label{A38}
\end{equation}
The functional version is understood with a regularization prescription compatible with the conjugation pairing, as discussed below.

\subsection{Standard complex Gaussian as two real Gaussians}
It is useful to recall the standard result. Consider one complex variable,
\begin{equation*}
    z=x+iy ,
\end{equation*}
and a positive real number \(a>0\). Then,
\begin{equation*}
    z^{\ast} a z = a (x^2+y^2). 
\end{equation*}
Therefore,
\begin{equation*}
    \int_{C} dz d z^{\ast} e^{-a z^{\ast} z}
\end{equation*}
is, up to the normalization of the complex measure,
\begin{equation}
    \int_{-\infty}^{\infty} dx \int_{-\infty}^{\infty} dy e^{-a(x^2 + y^2)}.
\label{A39}
\end{equation}
Hence,
\begin{equation}
     \int_{C} d^2 z e^{-a z^{\ast} z} = \frac{\pi}{a}.
\label{A40}
\end{equation}
For \(N\) complex variables and a Hermitian positive-definite matrix \(A\),
\begin{equation}
    \int_{\mathbb{C}^N} d^{2N} z e^{-z^{\dagger} A z} = \frac{\pi^N}{\det A}.
\label{A41}
\end{equation}
Writing,
\begin{equation*}
    A = A_R + i A_I,
\end{equation*}
its real representation is,
\begin{equation}
    \Re(A) =  \begin{pmatrix}
A_{R} & -A_{I}  \\
A_{I} & A_{R} \\
\end{pmatrix}.
\label{A42}
\end{equation}
Then,
\begin{equation}
z^{\dagger} A z = \begin{pmatrix}
x \\
y
\end{pmatrix}^{T}
\Re(A)
\begin{pmatrix}
x \\
y
\end{pmatrix} ,
\label{A43}
\end{equation}
and,
\begin{equation}
    \det \Re(A) = |\det A |^2.
\label{A44}
\end{equation}
This proves the complete equivalence between the ordinary complex Gaussian and a Gaussian over twice as many real variables.

\subsection{Off-diagonal Gaussian structure of the projected hypercomplex sector}
The projected action is not generically of the standard form:
\(\overline{\Omega^{+}} \mathcal{D} \Omega^{+}\). Instead,
\begin{equation*}
    S_{+} = \overline{\Omega^{-}} \mathcal{D}_{+} \Omega^{+}.
\end{equation*}
That is, it couples two different complex fields. This means that the real representation is an off-diagonal Gaussian. For a single mode, write,
\begin{equation*}
    u = \begin{pmatrix}
\Re(\Omega^{+}) \\
\Im (\Omega ^{+})
\end{pmatrix}, \hspace{1cm}
v = \begin{pmatrix}
\Re(\Omega^{-}) \\
\Im (\Omega ^{-})
\end{pmatrix}.
\end{equation*}
Then,
\begin{equation}
    S_{+} = v^{T} \textup{Q}_{+} u,
\label{A45}
\end{equation}
with a \(2 \times 2\) complex or realified operator \(\textup{Q}_{+}\). This can always be written as,
\begin{equation}
 S_{+} = \frac{1}{2} \begin{pmatrix}
u \\ v
\end{pmatrix}^{T}
\begin{pmatrix}
0 & \textup{Q}_{+}^{T} \\
\textup{Q}_{+} & 0 \\
\end{pmatrix} 
\begin{pmatrix}
u \\v
\end{pmatrix}.
\label{A46}
\end{equation}
Therefore, it is Gaussian over real variables. To illustrate the origin of the indefinite directions independently of the complex phase structure, consider first a real nonsingular representative \(\textup{Q}\). The symmetric matrix,
\begin{equation*}
     \begin{pmatrix}
0 & \textup{Q}^{T} \\
\textup{Q} & 0 \\
\end{pmatrix}
\end{equation*}
has, for real nonsingular \(\textup{Q}\), eigenvalues appearing in opposite-sign pairs. In the simplest scalar case, \(\textup{Q} = q > 0\), 
one obtains,
\begin{equation*}
     \begin{pmatrix}
0 & q \\
q & 0 \\
\end{pmatrix},
\end{equation*}
with eigenvalues \(+q, -q\) . Hence the projected form is not positive definite on the naive real contour.

\subsection{Diagonalizing the off-diagonal real Gaussian}
For this illustrative real symmetric representative, the structure can be made even clearer by defining,
\begin{equation}
    r = \frac{u+v}{\sqrt{2}} , \hspace{1.5cm}   
    s = \frac{u-v}{\sqrt{2}}.
\label{A47}
\end{equation}
For a symmetric \(\textup{Q}\),
\begin{equation}
    v^{T} \textup{Q} u = \frac{1}{2} r^{T} \textup{Q} r - \frac{1}{2} s^{T} \textup{Q} s.
\label{A48}
\end{equation}
Thus,
\begin{equation}
    S_{+} = \frac{1}{2}  r^{T} \textup{Q} r - \frac{1}{2} s^{T} \textup{Q} s.
\end{equation}
The Gaussian nature is explicit, but so is its indefiniteness. If \(\textup{Q} > 0\), the \(r\) directions are damped whereas the \(s\) directions grow on the naive real contour. Therefore,
\begin{equation*}
    \int_{\mathbb{R}^{2N}} dr \, ds \, e^{-S_{+}}
\end{equation*}
is not convergent as an ordinary Euclidean integral.

\subsection{Steepest-descent convergence prescription}
A convergent Gaussian may nevertheless be defined by choosing the steepest-descent contour for the unstable directions. For the variables \(s\), perform the contour rotation,
\begin{equation}
    s= i \widetilde{s}, \hspace{1.5cm}  \widetilde{s} \in \mathbb{R}^{N}.
\label{A50}
\end{equation}
Then,
\begin{equation*}
    - \frac{1}{2} s^{T} \textup{Q} s \qquad \rightarrow \qquad  + \frac{1}{s} \widetilde{s}^{T} \textup{Q} \widetilde{s}.
\end{equation*}
Consequently,
\begin{equation}
    S_{+} \qquad \rightarrow \qquad \frac{1}{2} r^{T} \textup{Q} r + \frac{1}{2} \widetilde{s}^{T} \textup{Q} \widetilde{s}.
\label{A51}
\end{equation}
For \(\textup{Q} > 0\), this is positive definite. 
Thus, for this illustrative real representative, the Gaussian becomes convergent after an appropriate deformation of the integration cycle. This construction motivates the phase-dependent contour rotation used for the actual complex modal problem below. For this illustrative case, the corresponding contour is,
\begin{equation}
    \mathcal{C}_{+} : \hspace{1cm} r \in \mathbb{R}^{N} , \qquad  s \in i \mathbb{R}^{N}.
\label{A52}
\end{equation}
The \(J^{-}\) sector is assigned the conjugate contour,
\begin{equation}
    \mathcal{C}_{-} = \mathcal{C}_{+}^{\ast}.
\label{A53}
\end{equation}
This ensures \(Z_{-} = Z_{+}^{\ast}\).

\subsection{Infinite-dimensional interpretation}
The functional integral is defined mode by mode. For the determinant and steepest-descent analysis we now return, as stated after Eq. (\ref{A24}), to the independent bilinear complex coordinates of the projected theory. The passage from the real-component representation (\ref{A25}) to the reduced bilinear modal form does not amount to taking the determinant of \(\mathbb{K}_{+}\). Rather, the projected functional measure is evaluated on the middle-dimensional integration cycle determined by the independent bilinear modal coordinates. The induced modal measure and its Jacobian must therefore be understood together with the contour prescription. With the induced modal measure on this contour, each eigenvalue \(\lambda_{\alpha}\) of \(\mathcal{D}_s\) contributes one complex bosonic Gaussian factor. After the Fourier-Matsubara decomposition, each independent mode therefore takes the reduced form,
\begin{equation}
    S_{\alpha} = \lambda_{\alpha} v_{\alpha} u_{\alpha},  
\label{A54.1}
\end{equation}
where
\begin{equation}
    \lambda_{\alpha} \equiv \lambda_{s} (n, \textbf{p}),
\label{A54.2}
\end{equation}
is an eigenvalue of the Euclidean operator \(\mathcal{D}_{s}\). Although the real-component representation involves four real field components, these integrations should be understood jointly as the real representation of the projected complex Gaussian measure, rather than as four independent copies of the same scalar determinant. Consequently, the determinant power is determined by the complete bilinear modal block. The reduced modal representation used below reproduces the inverse determinant associated with the corresponding complex bosonic mode.

\vspace{0.2cm}
\noindent
Introducing,
\begin{equation}
    r_{\alpha} = \frac{u_{\alpha} + v_{\alpha}}{\sqrt{2}},  \hspace{1.5cm}
    s_{\alpha} = \frac{u_{\alpha} - v_{\alpha}}{\sqrt{2}},
\end{equation}
gives,
\begin{equation}
    S_{\alpha} = \frac{\lambda_{\alpha}}{2} (r_{\alpha}^2 - s_{\alpha}^2).
\label{A55}
\end{equation}
The convergent cycle is obtained by selecting the directions in which,
\begin{equation*}
    Re(\lambda_{\alpha} r^2_{\alpha}) > 0,
\end{equation*}
and
\begin{equation*}
      Re(-\lambda_{\alpha} s^2_{\alpha}) > 0.
\end{equation*}
Equivalently, each integration direction is rotated to a steepest-descent contour determined by the phase of \(\lambda_{\alpha}\). This is the infinite-dimensional analogue of the standard analytic continuation used in complex Gaussian integration.

\subsection{Determinant obtained from the real Gaussian}
After the contour is chosen, each complex mode produces the expected determinant. For a standard complex bosonic mode \(Z_{\lambda} \propto \frac{1}{\lambda} \). Its realified pair has determinant \(|\lambda|^2\). Hence,
\begin{equation*}
    (\det R_{\lambda})^{-1/2} = \frac{1}{|\lambda|}.
\end{equation*}
The missing phase is supplied by the orientation/analytic continuation of the integration contour. Thus, for the two conjugate sectors,
\begin{equation}
    Z_{+} \propto (\Det \mathcal{D}_{+})^{-1} , \hspace{1cm}
    Z_{-} \propto (\Det \mathcal{D}_{-})^{-1} = \left([\Det \mathcal{D}_{+}]^{-1} \right)^{\ast}.
\label{A57}
\end{equation}
Therefore,
\begin{equation}
    \ln Z_{\mathbb{H}} = - J^{+} \Tr_{\texttt{reg}} \ln \mathcal{D}_{+} - J^{-} \Tr_{\texttt{reg}} \ln \mathcal{D}_{-}.
\label{A58}
\end{equation}
The regularized trace is understood using the same principal logarithmic branch fixed in Sec. \ref{seccion 3.5}, and the regularization prescription is chosen to preserve the conjugation pairing of the two idempotent spectra. In particular,
\begin{equation*}
    \Tr_{\texttt{reg}} \ln \mathcal{D}_{-} = \left[\Tr_{\texttt{reg}} \ln \mathcal{D}_{+} \right]^{\ast}.
\end{equation*}
This is precisely the determinant structure used in the Matsubara derivation of the main text. 
Equivalently, since \(\mathbb{C}^2 \simeq \mathbb{R}^4\), the projected complex functional measure admits the corresponding real-component representation: 
\begin{equation}
    \mathcal{D} \Omega^{+} \mathcal{D} \overline{\Omega^{+}} \mathcal{D} \Omega^{-} \mathcal{D} \overline{\Omega^{-}} \leftrightarrow   \mathcal{D} \phi_1   \mathcal{D} \phi_2  \mathcal{D} \chi_1  \mathcal{D} \chi_2.
\label{A59}
\end{equation}
The exponent remains quadratic, and consequently the functional integral remains Gaussian. However, Gaussian character does not imply positive definiteness. A positive Euclidean Gaussian is obtained immediately only when the quadratic form has the ordinary sesquilinear structure \(\phi^{\dagger} A \phi\), \(A >0\). The projected idempotent sectors instead possess an off-diagonal bilinear structure; thus their convergence requires a specified complex integration cycle.

\vspace{0.2cm}
\noindent
This feature is not an inconsistency of the hypercomplex theory. The two projected contributions are related through conjugation: \(S_{-} = S_{+}^{\ast} \), and hence \(Z_{-} = Z_{+}^{\ast} \). Their recombination gives \(Z_{\mathbb{H}} = J^{+} Z_{+} + J^{-} Z_{+}^{\ast} \), which is Hermitian in the hypercomplex sense. The two idempotent components should therefore not be interpreted as two unrelated real Euclidean theories. They are conjugate projections of a single hypercomplex functional integral.

\subsection{Convergence criterion after Fourier decomposition}
For each mode,
\begin{equation*}
    \lambda_{\pm} (n, \textbf{p}) = \left(\omega_{n} - i \mu_{\pm} \right)^2 + E^2_{\textbf{p}.}
\end{equation*}
A singular Gaussian occurs when \(\lambda_{\pm} = 0\). Away from such zeros, a steepest-descent cycle exists locally for every nondegenerate quadratic mode. Thus the basic nondegeneracy condition is,
\begin{equation}
    \lambda_{\pm} (n, \mathbf{p}) \neq 0,
\label{A61}
\end{equation}
for every retained regulated mode. For the thermal, non-condensed sector, the sufficient physical conditions used in the main text are,
\begin{equation}
    m^2_{\texttt{eff}} > 0,
\label{A61}
\end{equation} 
and
\begin{equation}
    |\mu| < m_{\texttt{eff}}.
\label{A63}
\end{equation}
These conditions guarantee a real quasiparticle spectrum and convergence of the thermal occupation expansion. These conditions do not imply positive definiteness of the naive \(J^{\pm}\) real quadratic form.

\subsection{Steepest-descent prescription and its applicability to the projected Gaussian functional integral}

The steepest-descent method provides a prescription for defining oscillatory or complex-valued integrals by extending the integration variables to the complex domain and selecting integration cycles along which the real part of the exponent produces exponential damping. In its modern formulation, this construction is naturally described in terms of Morse or Picard–Lefschetz theory, where the original integration domain is replaced, or analytically continued, by suitable middle-dimensional cycles associated with critical points of the complexified action \cite{witten}. The same framework has been applied to complex path integrals in quantum field theory, including scalar field theories at finite chemical potential \cite{cristoforetti,cristoforetti 2}.  To state the prescription more precisely, consider first a finite-dimensional integral
\begin{equation*}
    I = \int_{\mathcal{C}} d^{N} z e^{-S(z)},
\end{equation*}
where \(S(z)\) is holomorphic in the complexified variables \(z^a\). The relevant critical points \(z_{\sigma}\) are determined by,
\begin{equation*}
 \left.\begin{matrix}
 \frac{\partial S}{\partial z^a}
\end{matrix}\right|_{z=z_{\sigma}} = 0.
\end{equation*}
A steepest-descent cycle associated with \(z_{\sigma}\) is a real \(N\)-dimensional integration manifold in the \(N\)-complex-dimensional configuration space. It may be characterized through the gradient flow of the Morse function,
\begin{equation*}
    h(z, \overline{z}) = \text{Re} S(z).
\end{equation*}
Using an outward flow convention,
\begin{equation*}
    \frac{dz^a}{dt} = \frac{\overline{\partial S}}{\partial z^a},
\end{equation*}
one obtains
\begin{equation*}
    \frac{d}{dt} \text{Re} S = \sum_{a} \left | \frac{\partial S}{\partial z^a}\right | ^2 \geq 0,
\end{equation*}
whereas \(\frac{d}{dt} \text{Im} S = 0.\)
Thus, along the steepest-descent directions, the real part of the action increases monotonically while its imaginary part remains constant. Then,
\begin{equation*}
    \text{Re} \, S \quad \rightarrow \quad \infty
\end{equation*}
at the asymptotic ends of a convergent cycle and therefore,
\begin{equation*}
    e^{-S} \quad \rightarrow \quad 0
\end{equation*}
exponentially. Equivalently, one may define the same cycle using the opposite sign in the flow equation and regard it as the set of trajectories that approach the critical point under downward flow. These are equivalent conventions for describing the corresponding Lefschetz thimble. This gradient-flow construction and the constancy of the imaginary part of the action are central ingredients of the Picard–Lefschetz formulation of complex path integrals \cite{witten,cristoforetti 2}.

\vspace{0.3cm}
\noindent
For the present field theory, this prescription must first be understood with a regulator. We therefore consider the system at finite spatial volume and introduce a spectral or ultraviolet cutoff, hence that only a finite number of Fourier-Matsubara modes is retained. The functional integral then becomes an ordinary finite-dimensional complex Gaussian integral. The continuum theory is recovered only after the mode-by-mode integration has been defined and the corresponding infinite-dimensional determinant has been regularized. After the Fourier-Matsubara decomposition, the quadratic operators of the two idempotent sectors have eigenvalues,
\begin{equation*}
    \lambda_s (n, \textbf{p}) = \left(\omega_{n} - i \mu_s \right)^2  +  E_{\textbf{p}}^2 , \hspace{1.5cm} s=\pm,
\end{equation*}
where
\begin{equation*}
    \mu_s = \mu - i s \frac{\gamma}{2} , \hspace{1cm}   E_{\textbf{p}}^2 =\textbf{p}^2 + m^2_{\texttt{eff}} , \hspace{1cm}    m^2_{\texttt{eff}} = m^2 - \frac{\gamma^2}{4}.
\end{equation*}
Equivalently,
\begin{equation*}
    \lambda_s  (n, \textbf{p}) = \left(\omega_{n} - s \frac{\gamma}{2} - i \mu \right)^2  +  E_{\textbf{p}}^2 .
\end{equation*}
Let
\begin{equation*}
    \alpha \equiv (s, n , \textbf{p})
\end{equation*}
denote collectively the sector and mode labels. For each mode, the projected quadratic action can be written schematically as,
\begin{equation*}
    S_{\alpha} = \lambda_\alpha x_\alpha y_{\alpha}.
\end{equation*}
The corresponding stationary-point equations are,
\begin{equation*}
    \frac{\partial S_\alpha}{\partial x_{\alpha}} = \lambda_{\alpha} y _{\alpha} = 0 ,   \hspace{1.5cm}  
    \frac{\partial S_\alpha}{\partial y_{\alpha}} = \lambda_{\alpha} x_{\alpha} = 0.
\end{equation*}
For \(\lambda_\alpha \neq 0\), the origin, \(x_{\alpha} = y_{\alpha} = 0\), is the unique critical point. Its Hessian is,
\begin{equation*}
    H_{\alpha} = \begin{pmatrix}
0 & \lambda _\alpha  \\
\lambda _\alpha  & 0  \\
\end{pmatrix},
\end{equation*}
with,
\begin{equation*}
    \det H_{\alpha} = - \lambda^{2}_{\alpha} \neq 0.
\end{equation*}
Therefore, every mode with \(\lambda_{\alpha} \neq 0\) corresponds to a non-degenerate Gaussian saddle. This non-degeneracy is essential: if an eigenvalue vanishes, the quadratic saddle develops a flat direction and cannot be included in the ordinary Gaussian determinant. The steepest-descent cycle can be constructed explicitly in the present case. Write,
\begin{equation*}
    \lambda_{\alpha} = |\lambda_{\alpha}| e^{i \vartheta_{\alpha}},
\end{equation*} 
and introduce
\begin{equation*}
    r_{\alpha} = \frac{x_\alpha + y_\alpha}{\sqrt{2}} , \hspace{1 cm}
    q_{\alpha} = \frac{x_\alpha - y_\alpha}{\sqrt{2}}.
\end{equation*}
The modal action becomes,
\begin{equation}
    S_{\alpha} = \frac{\lambda_{\alpha}}{2} \left(r^2_{\alpha} - q^2_{\alpha} \right).
\end{equation}
If both variables were integrated along their naive real directions, the two quadratic terms would have opposite signs, and the resulting Gaussian would not be convergent. Therefore, the projected functional integral should not be interpreted as an ordinary integral over the naive real contour. Instead, define the middle-dimensional complex cycle by
\begin{equation*}
    r_{\alpha} = e^{- i \vartheta_{\alpha}/2} \, u_{\alpha}, \hspace{1.2cm}
    q_{\alpha} = i e^{-i \vartheta_{\alpha}/2} \, v_{\alpha}, \hspace{1.2cm}
    u_{\alpha}, v_{\alpha} \in \mathbb{R},
\end{equation*}
the substitution gives,
\begin{equation*}
    S_{\alpha} = \frac{|\lambda_{\alpha}|}{2} \left( u^2_{\alpha} + v^2_{\alpha} \right).
\end{equation*}
Hence,
\begin{equation*}
    \text{Re} \, S_{\alpha} = \frac{|\lambda_{\alpha}|}{2}  \left( u^2_{\alpha} + v^2_{\alpha} \right) \geq 0,
\end{equation*}
and, more importantly,
\begin{equation*}
        \text{Re} S_{\alpha} \quad \rightarrow \quad + \infty, \hspace{1 cm}  \text{as} \hspace{1cm}
        \left( u^2_{\alpha} + v^2_{\alpha} \right) \quad \rightarrow \quad \infty. 
\end{equation*}
It follows that,
\begin{equation*}
    \left | e^{-S_{\alpha}}\right | = \exp \left[- \frac{|\lambda_{\alpha}|}{2} \left( u^2_{\alpha} + v^2_{\alpha} \right)    \right] \quad \rightarrow \quad 0
\end{equation*}
exponentially at the asymptotic ends of the integration cycle. Therefore, the Gaussian integral associated with every non-degenerate mode is absolutely convergent on this cycle. This construction shows explicitly why the steepest-descent prescription is applicable to the present theory. It does not rely merely on assuming the existence of a suitable contour: because the action is quadratic, the required integration cycle can be written analytically for every mode. Moreover, there is only one quadratic saddle for each nonzero eigenvalue; therefore, the complications associated with multiple competing saddles do not arise at this level. The physical thermal domain considered in the present work provides an additional consistency condition. The real part of the eigenvalues is,
\begin{equation*}
    \text{Re} \lambda_{s} (n, \textbf{p}) = \left(\omega_{n} - s \frac{\gamma}{2} \right)^2  +  E_{\textbf{p}}^2 - \mu^2. 
\end{equation*}
Under the conditions
\begin{equation*}
    m^2_{\texttt{eff}} > 0, \hspace{1.5cm} |\mu| <  m_{\texttt{eff}},
\end{equation*}
one has
\begin{equation*}
    E^2_{\textbf{p}} - \mu^2 \geq m^2_{\texttt{eff}} - \mu^2 > 0.
\end{equation*}
Thus,
\begin{equation*}
    \text{Re} \, \lambda_{s} (n, \textbf{p}) > 0
\end{equation*}
for all Matsubara frequencies and momenta in the thermal domain studied here. The spectrum therefore remains in the open right half of the complex plane. In particular, no eigenvalue crosses the origin within this domain, and its phase,
\begin{equation*}
  \vartheta_{\alpha} =   \arg{\lambda_{\alpha}},
\end{equation*}
can be chosen continuously. The corresponding steepest-descent cycles can consequently also be selected continuously throughout this parameter region. The two idempotent sectors satisfy the conjugation relation,
\begin{equation*}
    \lambda_{-} (-n, \textbf{p}) = \lambda_{+} (n, \textbf{p})^{\ast}.
\end{equation*}
The associated integration cycles may therefore be chosen as conjugate cycles,
\begin{equation*}
    \mathcal{C}_{-} = \mathcal{C}_{+}^{\ast},
\end{equation*}
which is consistent with
\begin{equation*}
    Z_{-} = Z_{+}^{\ast}.
\end{equation*}
An additional feature of the contour prescription is that it preserves the complex phase of the modal determinant. The contour rotation contributes a complex Jacobian, whereas the positive Gaussian integral itself produces a factor proportional to \(\frac{1}{|\lambda_{\alpha}|}\). Combining both contributions yields, up to an overall normalization and orientation independent of the dynamical parameters,
\begin{equation*}
    Z_{\alpha} \propto \frac{1}{\lambda_{\alpha}}.
\end{equation*}
Thus, the complex phase of the Gaussian determinant is not lost by the real-component representation; it is encoded in the orientation of the corresponding complex integration cycle. It is essential, however, to distinguish the convergence of each modal Gaussian from the definition of the complete functional determinant. The condition
\begin{equation*}
    \lambda_{\alpha} \neq 0,
\end{equation*}
ensures that the individual critical point is non-degenerate, while the steepest-descent prescription ensures convergence of the corresponding finite-dimensional Gaussian integral. Neither statement by itself implies convergence of the infinite product,
\begin{equation*}
    \Det \mathcal{D}_{s} = \prod_{n, \textbf{p}} \lambda_{s} (n, \textbf{p}).
\end{equation*}
The latter is a functional determinant and must be defined through a suitable regularization prescription. In the present finite-temperature calculation, this distinction becomes explicit after the Matsubara summation: the ultraviolet-divergent zero-point contribution is isolated in the vacuum term, whereas the thermal contribution is ultraviolet finite. If the parameters are analytically continued to a point at which an eigenvalue satisfies
\begin{equation*}
    \lambda_{\alpha} = 0,
\end{equation*}
the associated Gaussian saddle becomes degenerate and the preceding construction no longer applies to that mode. Such modes must therefore be separated from the Gaussian fluctuation determinant rather than simply discarded. We denote the determinant over the nonzero fluctuation spectrum by,
\begin{equation*}
    {\Det}' \mathcal{D}_s = \prod_{\lambda_{\alpha} \neq 0} \lambda_{\alpha},
\end{equation*}
and write schematically,
\begin{equation*}
Z_{s} = Z_{ker,s} \, \left[ {\Det}' \mathcal{D}_{s} \right]^{-1}_{\texttt{reg}}.
\end{equation*}
Here \(Z_{ker,s}\) denotes the contribution associated with the kernel of \(\mathcal{D}_s\) if true zero eigenvalues occur under analytic continuation of the parameters. It should not be confused with the static Fourier-mode contribution \(Z_{0,s}\) isolated in the main text. The prime indicates the exclusion of kernel modes from the Gaussian fluctuation determinant, whereas ‘reg’ emphasizes that the remaining infinite product must still be regularized.

\vspace{0.2cm}
\noindent
Finally, the present quadratic theory does not require the full machinery needed for a general Picard–Lefschetz decomposition. In interacting theories, several critical points may contribute, their associated thimbles may change across Stokes regions, and the original integration cycle may have to be expressed as a combination of several thimbles. Such issues are central to the more general application of Lefschetz-thimble methods to quantum field theory \cite{witten, cristoforetti, cristoforetti 2}. In the present Gaussian theory, by contrast, the relevant cycle is obtained explicitly mode by mode, and the analysis reduces to the non-degeneracy of the quadratic saddle, the asymptotic positivity of \(\text{Re} S_{\alpha}\), and the independent regularization of the resulting functional determinant.

\end{appendices}


\end{document}